\documentclass[aps,prd,onecolumn,showpacs,floatfix,preprintnumbers,amsmath,amssymb,nofootinbib]{revtex4}
 \usepackage{graphicx,color}

\def\fnl{{f_{\rm{NL}}}} 

\def\fnle{{\widehat \fnl}}

\def\VEV#1{\left\langle #1 \right\rangle}

\def\fnlnull{\fnle}

\def\wigner#1#2#3#4#5#6{ \left( \begin{array}{ccc} #1 & #3 & #5
\\ #2 & #4 & #6 \\ \end{array} \right)}

\newcommand{\beq}{\begin{equation}}
\newcommand{\eeq}{\end{equation}}
\newcommand{\beqa}{\begin{eqnarray}}
\newcommand{\eeqa}{\end{eqnarray}}

\newcommand{\Npix}{N_{\mathrm{pix}}}

\newcommand{\lmax}{{l_{\mathrm{max}}}}

\begin{document}
 \raggedbottom
\title{An improved estimator for non-Gaussianity in cosmic microwave background observations}
\author{Tristan L.~Smith$^{1}$, Daniel Grin$^{2}$, and Marc Kamionkowski$^3$}
\affiliation{$^{1}$Berkeley Center for Cosmological Physics \& Berkeley Lab
University of California, Berkeley, CA 94720, USA}
\affiliation{$^{2}$School of Natural Sciences, Institute for Advanced Study, Princeton, New Jersey 08540, USA}
\affiliation{$^{3}$Department of Physics and Astronomy, Johns Hopkins University, Baltimore, MD 21218, USA}
 \date{\today}

\begin{abstract}
An improved estimator for the amplitude $\fnl$ of local-type non-Gaussianity from the cosmic microwave background (CMB) bispectrum is discussed. The standard estimator is constructed to be optimal in the zero-signal (i.e., Gaussian) limit.  When applied to CMB maps which have a detectable level of non-Gaussianity the standard estimator is no longer optimal, possibly limiting the sensitivity of future observations to a non-Gaussian signal. 
Previous studies have proposed an improved estimator by using a realization-dependent normalization.  Under the approximations of a flat sky and a vanishingly thin last-scattering surface, these studies showed that the variance of this improved estimator can be significantly smaller than the variance of the standard estimator when applied to non-Gaussian CMB maps. 
 Here this technique is generalized to the full sky and to include the full radiation transfer function, yielding expressions for the improved estimator that can be directly applied to CMB maps. The ability of this estimator to reduce the variance as compared to the standard estimator in the face of a significant non-Gaussian signal is re-assessed using the full CMB transfer function. As a result of the late time integrated Sachs-Wolfe effect, the performance of the improved estimator is degraded. If CMB maps are first cleaned of the late-time ISW effect using a tracer of foreground structure, such as a galaxy survey or a measurement of CMB weak lensing, the new estimator \textit{does} remove a majority of the excess variance, allowing a higher significance detection of $\fnl$.
\end{abstract}
\maketitle
\section{Introduction}

Over the past two decades our understanding of the physics of the early universe 
has gone from speculative to precise.  We have numerous data sets which probe both the overall expansion 
of the universe as well as the statistics of the large-scale structure we see today.  In addition to these observations, 
our standard cosmological model, with only six free parameters, provides a good fit to all of these data sets (see, e.g.~\cite{Komatsu:2010fb}). The standard cosmological model relies on some basic assumptions about the origin and evolution of the universe: namely that soon after the big bang, the universe underwent a period of cosmic inflation during which nearly Gaussian perturbations were produced in an otherwise isotropic and homogeneous universe.  After this period the universe was `reheated'-- i.e., populated with a thermal plasma of standard model particles.  Particle physics dictates the interactions between the constituents of this primordial fluid, while general relativity dictates how the perturbations grow to form the structures we observe in both the clustering of galaxies, as well as in the anisotropies of the cosmic microwave background (CMB).  

With increasingly precise observations, we may test the foundations of the standard cosmological model. Here we will be concerned with testing the assumption that the statistics of CMB fluctuations are Gaussian.  Although small levels of non-Gaussianity may develop through non-linearities in the standard cosmological model \cite{Salopek:1990jq,1996PhRvD..53.2920P,Mollerach:1997up,Mollerach:1995sw,1995ApJ...454..552M}, significant departures from Gaussianity can only be explained by changes to the fundamental physics of the early universe. 
For example, a detection of primordial non-Gaussianity could yield information on the interactions of the field (or fields) that seed the primordial curvature perturbation. If this field is the one that drives inflation (the inflaton), the measurement could yield insight into the detailed  physics of inflation. If the curvature perturbation is seeded by a field that is sub-dominant during inflation (as in the curvaton scenario \cite{Mollerach:1990ue,Lyth:2002my,Lyth:2001nq,Gordon:2002gv}), the primordial fluctuations may also be significantly non-Gaussian. Some more exotic possibilities, such as a non-canonical kinetic term for the inflaton \cite{Cheung:2007st,Alishahiha:2004eh,ArkaniHamed:2003uz}, spatially modulated reheating \cite{Dvali:2003em}, and novel initial vacuum states for the fluctuations \cite{Martin:1999fa}, could also be probed by a detection or limit to non-Gaussianity in the CMB. 

There are an infinite number of ways in which the statistics of the CMB may be non-Gaussian, although the effective field theory approach shows that inflationary theories can only produce 
three forms for the non-Gaussian primordial three-point correlation function \cite{Cheung:2007st}.  In this study we restrict our attention to the `local model' of non-Gaussianity, in which the primordial curvature perturbation, $\Phi$, can be written in terms of an auxiliary Gaussian field $\phi$ as \cite{Salopek:1990jq,Gangui:1993tt,1994PhRvD..50.3684G,Komatsu:2001rj,Verde:1999ij}
\begin{equation}
 \Phi(\vec x) = \phi(\vec x) + \fnl \left[\phi^2(\vec x) - \VEV{\phi^2(\vec{x})}\right],\label{localansatz}
 \end{equation}
 where the amplitude $\fnl$ parameterizes the level of non-Gaussianity.  This model is particularly important because if $\fnl \neq 0$ were detected then \emph{all} single-field slow-roll inflation models would be ruled out \cite{Creminelli:2004yq,Smith:2011if}.  In addition, the local-type non-Gaussianity is predicted by the curvaton model \cite{Lyth:2002my,Lyth:2001nq,Gordon:2002gv}, and so a detection could provide support for, or constraints to, this model.

Although there are several ways to estimate the level of non-Gaussianity in the CMB, an estimate of the harmonic three-point function (known as the CMB bispectrum) is the most sensitive \cite{Komatsu:2001rj, Smith:2012ty}.  Given the large number of modes in the bispectrum (after restrictions due to statistical isotropy there are $l_{\rm max}^5$ terms in the bispectrum, where $\lmax(\simeq \sqrt{\Npix})$ is the maximum multipole measured by a given experiment) a full exploration of the likelihood surface is computationally prohibitive, and so attempts to constrain the level of non-Gaussianity are made through estimators which have been constructed to minimize variance under the null hypothesis-- i.e., when applied to CMB maps which are purely Gaussian \cite{Babich:2004yc,2005PhRvD..72d3003B}.  

A direct application of the minimum-variance null-hypothesis (MVNH) estimator of $f_{\rm NL}$ from the CMB bispectrum is also computationally expensive since the computation involves a very large number of  terms ($\sim l_{\rm max}^5$).  The Planck satellite \cite{Planck:2006aaT} will measure $l_{\rm max} \sim 2500$ multipole moments so a blind application of this estimator will take $\sim 10^{15}$ operations to compute!  The real computational expense is even higher, as thousands of simulations must be run to characterize the statistics of this estimator.  For a special set of non-Gaussian models (of which the local model is one) fast Fourier transforms (FFTs) may be used to greatly reduce this computational expense  \cite{Komatsu:2003iq, Smith:2006ud}. Using FFTs the estimator may then be evaluated with a computation time that scales as $l_{\rm max}^3$, considerably reducing the computational expense \cite{Komatsu:2003iq,Munshi:2009ik}. CMB data are currently consistent with vanishing non-Gaussianity \cite{Smith:2009jr,2008PhRvL.100r1301Y}; analysis of the WMAP 7-year data yields the 2-$\sigma$ constraint $-10<\fnl<74$ \cite{Komatsu:2010fb}. 

If future data remain consistent with the null hypothesis, then the standard MVNH estimator should have the minimum variance.  As the data improves, if $\fnl$ is found to be significantly different from zero, then the standard MVNH estimator is non-optimal and there may be other estimators which have a smaller variance (and hence increased signal-to-noise, $S/N$).  This is because 
the MVNH estimator was \emph{constructed} to have the minimum variance when $\fnl = 0$ and when applied to data with $\fnl \neq 0$ the variance is no longer minimized.  In particular, when applied to the local model, for $\fnl \neq 0$ the MVNH estimator exhibits a variance which is proportional to $f_{\rm NL}^{2}$, leading to a saturation of the $S/N$ of the estimator, even for large values of $\lmax$ \cite{Creminelli:2006gc, Liguori:2007sj,Smith:2011rm}.  This indicates that a new method to estimate $\fnl$ may extract a higher $S/N$ from the measurements. 

In one approach, estimators are abandoned, and the full likelihood surface is explored using Bayesian methods \cite{Elsner:2010hb}.  Although this approach has been shown to give improved constraints on $\fnl$, it is computationally expensive, taking over 150,000 CPU hours to compute a best fit-value and confidence interval for $\fnl$ with $\lmax = 512$.  In another approach, an improved estimator is constructed by defining a new realization-normalized estimator (RNE) \cite{Creminelli:2006gc, Smith:2011rm}.  This method normalizes the MVNH estimator using a particular combination of observed multipoles which are chosen so as to divide out the extra variance present.  
In Ref.~\cite{Creminelli:2006gc} the RNE was derived in the flat sky and Sachs-Wolfe limit (in which the temperature anisotropies are given by fluctuations in the gravitational potential at the surface of last scattering, $\Delta T/T = - \Phi/3$ \cite{White:1997vi}).  In these limits, it was shown in Ref.~\cite{Creminelli:2006gc} that the RNE successfully removes all the variance proportional to $f_{\rm NL}^{2}$ of the MVNH estimator when $\fnl \neq 0$.

In this work we generalize the RNE to include the full radiation transfer function and provide expressions that are valid on all-sky CMB maps. We compute the variance of the RNE and find that, unlike in the Sachs-Wolfe limit, only $\sim 50\%$ of the $\fnl$-dependent variance is removed. 

The residual $\mathcal{O}(f_{\rm NL}^{2})$ variance is a result of the late-time, large-scale integrated Sachs-Wolfe (ISW) effect, in which CMB anisotropies are generated by gravitational potential decay along the line-of-sight \cite{1968Natur.217..511R,1996PhRvL..76..575C,Boughn:2004zm,2006PhRvD..74f3520G,2008PhRvD..78d3519H,2008PhRvD..77l3520G,2010MNRAS.406....2F}. 
 To understand why the late-time ISW effect reduces our ability to remove this extra variance, we note that the form of the local-model non-Gaussianities implies that the bispectrum signal is dominated by harmonic-space triangles which correlate one large-scale mode with two small scale modes (i.e., `squeezed' triangles).  If the large-scale anisotropies are generated only by the Sachs-Wolfe effect then they can be inverted to give a direct measurement of the value of the large-scale gravitational potential at the surface of last scattering.  These measurements, in turn, allow us to directly infer the non-Gaussian contribution to the large-scale anisotropies, leading to a complete removal of all of the excess variance when $\fnl \neq 0$.  However, in the presence of \emph{both} the late-time ISW and Sachs-Wolfe effect, this inversion is not possible, leading to a degradation of the performance of the RNE.  We find, however, that if a tracer of foreground structure (such as a wide-field galaxy survey or the deflection field responsible for weak lensing of the CMB) is first applied to `clean' a CMB map of the late-time ISW effect then $\sim80-90\%$ of the $\fnl$-dependent variance may be removed by using the RNE.

We begin by summarizing the non-Gaussian local model and its bispectrum in Sec.~\ref{localintro}. We provide an intuitive, geometric argument for the origin of the $\fnl$-dependent variance that afflicts the local-model MVNH estimator in Sec.~\ref{sec:origin}. The realization-dependent normalization (RNE) is derived in Sec.~\ref{rdn} using a method that applies in the presence of the full radiation transfer function. In Sec.~\ref{varcalc} we compute the variance of the RNE using the full radiation transfer function, finding that the $\fnl$-dependent variance of the MVNH estimator is only partially removed. In Sec.~\ref{iswsub}, we show that the $\fnl$-dependent variance is further reduced by first cleaning the CMB of the late-time ISW contribution using a large-scale structure survey [such as the NRAO VLA Sky Survey (NVSS), or the next-generation Joint Dark Energy Mission (WFIRST) proposal] or a measurement of CMB lensing. Conventions for flat-sky approximations employed in numerical calculations are stated in Appendix \ref{flatskyappendix}. Detailed expressions needed to obtain the variance are derived in Appendix \ref{variance_details}.  In Appendix C we present a computationally efficient algorithm (using fast Fourier transforms) to compute our estimator on full-sky CMB maps. 

\section{The non-Gaussian local model}
\label{localintro}

First we will review the basic equations that relate to the definition and estimation of the CMB bispectrum, restricting attention to the local model as defined in Eq.~(\ref{localansatz}).
The CMB bispectrum is defined by
\begin{equation}
 B^{m_1 m_2 m_3}_{l_1 l_2 l_3} \equiv \VEV{a_{l_1 m_1}a_{l_2 m_2}a_{l_3 m_3}} = \mathcal{G}_{l_1 l_2 l_3}^{m_1 m_2 m_3} b_{l_1 l_2 l_3}, 
 \end{equation} 
 where the $a_{lm}$ are the usual multipole moments of the temperature map, $b_{l_1 l_2 l_3}$ is the reduced CMB bispectrum, and 
  the Gaunt integral is given by
\begin{eqnarray}
\mathcal{G}_{l_1 l_2 l_3}^{m_1 m_2 m_3} \equiv \int d^2 \hat{n} Y_{l_1 m_1}(\hat n) Y_{l_2 m_2}(\hat n) Y_{l_3 m_3}(\hat n) =\sqrt{\frac{\left(2l_1+1\right)\left(2l_{2}+1\right)\left(2l_{3}+1\right)}{4\pi}}\wigner{l_1}{0}{l_{2}}{0}{l_{2}}{0} \wigner{l_{1}}{m_{1}}{l_{2}}{m_{2}}{l_{3}}{m_{3}}\label{gaunt}
\end{eqnarray}
and $\wigner{l_{1}}{m_{1}}{l_{2}}{m_{2}}{l_{3}}{m_{3}}$ is a Wigner 3$J$-coefficient. 
A product of three multipoles  form an unbiased estimator for the angle-averaged CMB bispectrum:
\begin{equation}
B^{\rm obs}_{l_1 l_2 l_3} = \sum_{m_1 m_2 m_3} \wigner{l_{1}}{m_{1}}{l_{2}}{m_{2}}{l_{3}}{m_{3}}a_{l_1 m_1}a_{l_2 m_2}a_{l_3 m_3}
\end{equation}

The minimum-variance null-hypothesis (MVNH) estimator for $f_{\rm NL}$ is given by \cite{Babich:2004yc,2005PhRvD..72d3003B}
\begin{eqnarray}
\fnle = \sigma_{0}^{-2}\sum_{l_{1}\leq l_{2}\leq l_{3}}\frac{B_{l_{1}l_{2}l_{3}}^{\rm obs}B_{l_{1}l_{2}l_{3}}}{C_{l_{1}}C_{l_{2}}C_{l_{3}}},
\label{eq:biestimator}
\end{eqnarray}
where 
\begin{equation}
\left \langle a_{lm}a_{l'm'}^{*}\right \rangle=C_{l}\delta_{ll'}\delta_{mm'},
\end{equation}
and $B_{l_{1}l_{2}l_{3}}$ is the primordial (i.e., theoretical) angle-averaged bispectrum.
The normalization of this estimator is given by the variance under the null hypothesis, $\sigma_0^2$
 \cite{2008ApJ...678..578Y}:
\begin{eqnarray}
\sigma_{0}^{2}=\sum_{l_{1}\leq l_{2}\leq l_3} \frac{B_{l_{1}l_{2}l_{3}}^{2}}{\Delta_{l_{1}l_{2}l_{3}}C_{l_{1}}C_{l_{1}}C_{l_{3}}},
 \end{eqnarray}where $\Delta_{l_{1}l_{2}l_{3}}=1$ if $l_{1}\neq l_{2}\neq l_{3}$, $6$ if $l_{1}=l_{2}=l_{3}$, and $2$ otherwise.
Finally, the angle-averaged primordial bispectrum, $B_{l_1 l_2 l_3}$, is given in terms of the reduced bispectrum 
\begin{equation}
 B_{l_1l_2l_3}=\sqrt{\frac{\left(2l_1+1\right)\left(2l_{2}+1\right)\left(2l_{3}+1\right)}{4\pi}}\wigner{l_1}{0}{l_{2}}{0}{l_{3}}{0} b_{l_1 l_2 l_3}.
 \end{equation}

Now restricting attention to the local-model bispectrum, at any radial location along the line of sight  the primordial curvature potential, $\Phi$, can be decomposed into spherical harmonics. The local model ansatz in Eq.~(\ref{localansatz}) then implies that
\begin{equation}
\Phi_{l_1m_1}(r) = \phi_{l_1m_1}(r) + \fnl (-1)^{m_1} \sum_{l_{2}l_{3}}\sum_{m_{2}m_{3}}\mathcal{G}_{l_1 l_2 l_3}^{-m_1 m_2 m_3}\phi_{l_{2}m_{2}}(r)\phi_{l_{3}m_{3}}(r).
\label{localansatz_harmonic}
\end{equation}
This allows us to write the reduced bispectrum in a line-of-sight integral:
\begin{equation}
b_{l_1 l_2 l_3} = 2\left(\int r^2 dr \alpha_{l_1}(r) \beta_{l_2}(r) \beta_{l_3}(r) + {\rm cyclic~permutations.}\right),
\label{eq:redbi}
\end{equation}
 where the two filter-functions are given in terms of the transfer functions
 \begin{eqnarray}
 \alpha_l(r) &\equiv& \frac{2}{\pi}\int k^2dk j_l(kr) S_l(k), \label{eq:alpha}\\
 \beta_l(r) &\equiv& \frac{2}{\pi}\int k^2 dk j_l(kr) S_l(k) P(k),\label{eq:beta}
 \end{eqnarray}
 and $S_l(k)$ is the CMB temperature anisotropy transfer function \cite{Seljak:1996is}. 
The generalization to polarization is straightforward \cite{2008ApJ...678..578Y}; in this work we limit our attention to the signature of primordial non-Gaussianity in the CMB temperature.  

 Finally, we will need expressions for various auto and cross-correlations in terms of the distance along the 
 line-of-sight $r$.  The power-spectrum for the auxiliary Gaussian field $\phi$ is
\begin{equation}
\VEV{\phi(\vec k) \phi^*(\vec k')} = (2 \pi)^3 \delta^{(3)}(\vec k - \vec k') P(k),
\end{equation} 
where 
\begin{equation}
\phi(\vec x) =  \int \frac{d^3 \vec k}{(2\pi)^3} \phi(\vec k) e^{i \vec k \cdot \vec x}.
\end{equation}
These expressions lead to the line-of-sight autocorrelation \cite{Liguori:2003mb}
\begin{eqnarray}
\VEV{\phi_{lm}(r) \phi^*_{l' m'}(r')} &=& \frac{2}{\pi}\delta_{l,l'} \delta_{m,m'} \int k^2 dk P(k) j_l(kr) j_l(k r')\nonumber \\  &\equiv& \chi(r,r') \delta_{l,l'} \delta_{m,m'}.
\end{eqnarray}
Similarly, the temperature multipole moments can be written as \cite{Spergel:1999xn,Komatsu:2003iq}
 \begin{eqnarray}
a_{lm}=\int r^{2} dr \alpha_{l}(r) \Phi_{lm}(r).
\end{eqnarray}
With this we can write the line-of-sight cross-correlation 
\begin{eqnarray}
 \VEV{a_{lm} \phi^*_{l'm'}(r)} &=& \frac{2}{\pi} \delta_{l l'}\delta_{m m'}\int k^2 dk P(k)j_l(k r) S_l(k) \nonumber \\
 &\equiv& \delta_{l l'}\delta_{m m'} \beta_{l}(r).
 \end{eqnarray}
These expressions will be useful in the following discussion. 

\section{The origin of the increased variance \label{sec:origin}}

The standard non-Gaussian estimator written in Eq.~(\ref{eq:biestimator}) is only optimal under the null hypothesis: $\fnl = 0$.  In the case where $\fnl \neq 0$ the 
estimator becomes suboptimal (in the sense that it no longer saturates minimum variance attainable according to the Cramer-Rao bound, see Refs.~\cite{Babich:2004yc,Creminelli:2006gc}) and has a variance which depends on $\fnl$.  Specifically, in the flat-sky Sachs-Wolfe limit the variance scales as \cite{Creminelli:2006gc, Smith:2011rm,yd}
\begin{eqnarray}
\VEV{(\Delta \fnle)^2} &=& \sigma_0^2 + f_{\rm NL}^2\sigma_1^2 \nonumber,\\
&=& \frac{1}{72 A f_{\rm sky} l_{\rm max}^2 \ln(l_{\rm max})} +  f_{\rm NL}^2\frac{1}{2\ln^3(l_{\rm max})},
\label{eq:fullvar}
\end{eqnarray}
where $A$ is the amplitude of the power-spectrum of the primordial curvature perturbations, $P(k) = A k^{-3}$.
From this expression we can see that for $\fnl \neq 0$ that in the large $\lmax$ limit the variance flattens out and scales as $\ln^{-3}(\lmax)$.  

To understand the origin of this $\fnl$-dependent variance let us first consider a simple toy model.  Let $a_i$ 
 be a random variable with $\VEV{a_i} = 0$ and $\VEV{a_i a_j} = \sigma^2 \delta_{ij}$.  As we will discuss further, the $\fnl$-dependent variance 
 in the $\fnle$ estimator comes from terms which look like
 \begin{equation}
\hat{X} \equiv  \sum_{ij} a_i a_j.
\label{eq:toy}
 \end{equation}
The mean and variance of $\hat{X}$ are 
 \begin{eqnarray}
 \VEV{\hat{X}} &=& \sum_{ij} \VEV{a_i a_j} = N \sigma^2, \\
 \VEV{(\Delta \hat{X})^2} &=& \sum_{i j k l} \VEV{a_i a_j a_k a_l} - \sigma^4 \delta_{ij} \delta_{kl} = 2N^2 \sigma^4
 \end{eqnarray}
 where $N$ is the number of data points. 
 From this, it is clear that the $S/N$ of $\hat{X}$ is given by
 \begin{equation}
 \frac{S}{N} = \frac{\VEV{\hat{X}}}{\sqrt{\VEV{(\Delta \hat{X})^2}}} = \frac{1}{\sqrt{2}}.
 \end{equation}
 We can see that the $S/N$ is a constant.  The reason is simple: since the $a_i$ are 
 uncorrelated, the off diagonal terms in $\hat{X}$ with $i \neq j$ do not contribute to the 
 signal; on the other hand, they \emph{do} contribute to the noise, leading to a constant $S/N$\footnote{Of 
 course, if we were interested in estimating the variance we would choose a different 
 weighting than in Eq.~(\ref{eq:toy}) and use the estimator 
 $\sum_i a_i^2$ which has the same signal but a reduced variance compared 
 to Eq.~(\ref{eq:toy}).  As we will discuss further, the increased variance in the $\fnle$ estimator 
 cannot be reduced by choosing a new weighting.}.  We will see that something similar happens for the MVNH estimator.  
 
To understand the origin of the increased variance we expand the MVNH estimator in Eq.~(\ref{eq:biestimator}) in powers of $\fnl$
\begin{equation}
\fnle \approx \mathcal{B}_0 + \fnl \mathcal{B}_1,
\label{eq:fnl_expansion}
\end{equation}
with 
\begin{eqnarray}
     \mathcal{B}_0 &=&\sigma_{0}^{-2} \sum_{{l_{1}\leq l_{2}\leq l_{3}}}\sum_{m_{1}m_{2}m_{3}}
     \frac{a^{\rm L}_{l_1 m_1}a^{\rm L}_{l_2 m_2}a^{\rm L}_{l_3 m_3}}{C_{l_1} C_{l_2} C_{l_3} }B_{l_1l_2l_3}\wigner{l_{1}}{m_{1}}{l_{2}}{m_{2}}{l_{3}}{m_{3}},
     \\
        \mathcal{B}_1 &=&\sigma_{0}^{-2} \sum_{{l_{1}\leq l_{2}\leq l_{3}}}\sum_{m_{1}m_{2}m_{3}}
     \frac{a^{\rm NL}_{l_1 m_1}a^{\rm L}_{l_2 m_2}a^{\rm L}_{l_3 m_3}}{C_{l_1} C_{l_2} C_{l_3} }B_{l_1l_2l_3}\wigner{l_{1}}{m_{1}}{l_{2}}{m_{2}}{l_{3}}{m_{3}}+{\rm cyclic~permutations},
          \label{eq:B1}
\end{eqnarray}
where 
\begin{eqnarray}
a^{\rm L}_{lm}&=&\int r^{2} dr \alpha_{l}(r)\phi_{lm}(r),\\
a^{\rm NL}_{lm}&=& (-1)^m\sum_{l_1 l_2}\sum_{m_{1}m_{2}}\mathcal{G}^{-m m_1 m_2}_{l l_1 l_2}\int r^{2} dr \alpha_{l}(r)\phi_{l_{1}m_{1}}(r)\phi_{l_{2}m_{2}}(r). 
\label{anonlin_fullsky}
\end{eqnarray}
The total observed multipole moment is
\begin{equation}
a_{lm}\simeq a_{lm}^{\rm L}+\fnl a_{lm}^{\rm NL}.
\end{equation}

\begin{figure}[ht]
\begin{center}
\includegraphics[width=4in]{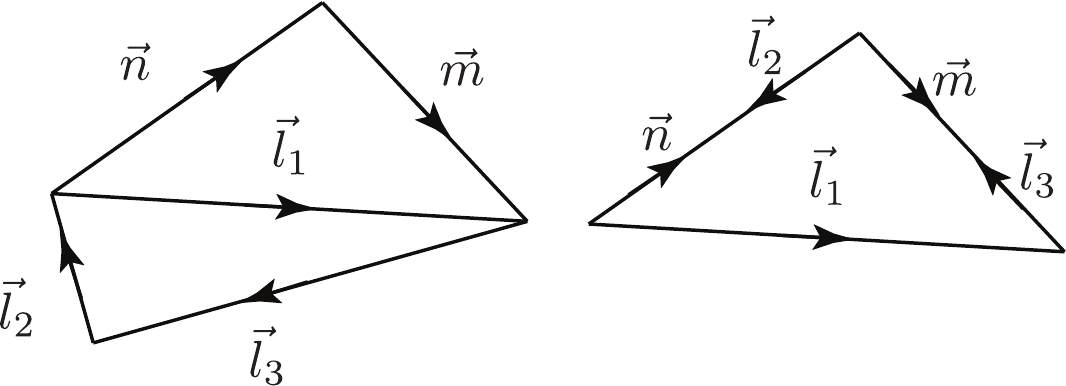}
\caption{A graphical representation of the temperature configurations that are included in the 
$\mathcal{B}_1$ part of the bispectrum estimator, Eq.~(\ref{eq:B1}). The full shape is composed of two distinct triangles.  
Only one triangle contributes to the mean, but all triangles contribute to the variance, leading to a $S/N$ which decreases 
slowly with $\lmax$. }
\label{fig:B1quad}
\end{center}
\end{figure}

We now shed light on the origin of the increased variance by considering the flat-sky and Sachs-Wolfe limit of Eq.~(\ref{eq:B1}).  In this limit the radiation transfer function is given by $S_l(k) = -j_l(k r_*)/3$ where $r_{*}$ is the conformal distance to the surface of last scattering.  We refer the reader to Appendix A for expressions which relate the full-sky to flat-sky expressions.  
In these limits we have \cite{Creminelli:2006gc,Smith:2011rm}:
\begin{eqnarray}
C_{l} &=& \frac{A}{9 \pi l(l+1)}, \\ 
\mathcal{B}_1 &=&3\sigma_{0}^2 \sum_{\vec m, \vec n, \vec l_2,\vec l_3} \frac{b_{l_1 l_2 l_3}}{2\Omega^3 C_{l_1} C_{l_2} C_{l_3}}\left\{ a_{\vec m} a_{\vec n}\delta_{\vec m + \vec n, \vec l_1}\right\} \left\{a_{\vec l_2} a_{\vec l_3}  \delta_{\vec l_2 + \vec l_3, - \vec l_1}\right\},
\label{eq:B1Tri}
\end{eqnarray}
where the primordial power-spectrum is taken to be scale invariant, $P(k) = A/k^3$. 
Rewriting $\mathcal{B}_1$ in this way we can see that it is a \emph{product of two triangles with a shared side $\vec l_1$}.  We show a 
graphical representation of this in the top section of Fig.~\ref{fig:B1quad}.
We can rewrite $\mathcal{B}_1$ schematically as 
\begin{equation}
\mathcal{B}_1 = \sum_{i,j} W(i) A_i A_j, 
\end{equation}
with $i$ labeling the triangle $i =\{\vec l_1,\vec l_2, \vec l_3\}$ and $j$ labeling the triangle $j = \{\vec l_1, \vec m, \vec n\} $. 

The mean is given by a sum over a single triangle, as shown in the right panel of Fig.~\ref{fig:B1quad}:
\begin{equation}
\VEV{\mathcal{B}_1} = \sum_{i,j} W(i)\VEV{A_i A_j} = \sum_{i} W(i) \VEV{A_i^2}.
\end{equation}
Therefore, we can see that the mean is composed of a sum of $(l_{\rm max}^2)^2=l_{\rm max}^4$ terms.  On the other hand, Fig.~\ref{fig:B1quad} shows that the variance of $\mathcal{B}_1$ 
is computed from a sum of a product of two triangles which gives $(l_{\rm max}^2)^4 = l_{\rm max}^8$.  As in the simple example given above, this shows that there are a large number of terms which do not contribute to the 
mean of $\mathcal{B}_1$ but \emph{do} contribute to the variance, leading to a signal-to-noise ($S/N$) which does not decrease as fast as one might have expected with increasing $\lmax$. 

It might seem that this can be corrected by choosing a different weight function $W(i)=W_{\vec l_1, \vec l_2, \vec l_3}$.  Since this weight was chosen to optimize the estimator under the null (i.e., $\fnl = 0$) hypothesis it is natural to think that a different weighting will be needed when $\fnl \neq 0$.  Unfortunately, since the terms in $\mathcal{B}_1$ which contribute to the 
variance but not the mean involve contractions between ($\vec l_1$,$\vec l_2$) and ($\vec n$,$\vec m$), it is clear that \emph{no choice of weight}, which depends only on the `observed' indices $(\vec l_1, \vec l_2, \vec l_3)$, will remove these terms.  Therefore, 
a new weighting cannot decrease the variance of this estimator in the high $S/N$ regime.  Another way to remove the terms which do not contribute to the 
signal but do contribute to the variance is to use a realization dependent normalization. 

Before we discuss how to construct a realization dependent normalization that reduces the variance in $\mathcal{B}_1$ let us first demonstrate in what sense the squeezed limit 
dominates the variance in $\mathcal{B}_1$.  
In the flat-sky Sachs-Wolfe limits, it has been shown that the variance is well approximated by \cite{Smith:2011rm}
\begin{equation}
 \VEV{(\Delta  \mathcal{B}_1)^2} \propto \sigma_0^4 \sum_{\vec l_1 + \vec l_2 + \vec l_3=0} \frac{b_{l_1 l_2 l_3}}{C_{l_1}} \sum_{\vec s_1 + \vec s_2 =\vec l_3} 
 \frac{b_{s_1 s_2 s_3}}{C_{s_1}} \delta_{\vec l_3 +
     \vec s_3,0}.
 \label{eq:lesscompact}
 \end{equation}
where $\{ \vec l\}$ indicates the sum is over $\vec l_1+\vec
l_2+\vec l_3=0$.  We can calculate this by first writing it in a less compact form:
In these limits the bispectrum is given by \cite{Babich:2004yc}
\begin{eqnarray}
b_{l_1 l_2 l_3}=
-6\lbrace C_{l_{2}}C_{l_{3}}+C_{l_{3}}C_{l_{1}}+C_{l_{1}}C_{l_{2}}\rbrace \label{swolfb}.
\end{eqnarray}
The second sum in Eq.~(\ref{eq:lesscompact}) is given by
\begin{equation}
 \sum_{\vec s_1 + \vec s_2 =\vec l_3} 
 \frac{b_{s_1 s_2 s_3}}{C_{s_1}} \delta_{\vec l_3 +
     \vec s_3,0}= -\frac{2}{3\pi}\sum_{\vec s_1 + \vec s_2 =\vec l_3} \left[\frac{s_{1}(s_{1}+1)}{s_{2}(s_{2}+1)l_{3}(l_{3}+1)}+\frac{1}{l_{3}(l_{3}+1)}+\frac{1}{s_{2}(s_{2}+1)} \right]
 \end{equation}
and the summand is dominated by triangles where one index is much less than the other two (i.e., the squeezed limit) so that 
 \begin{eqnarray}
 \sum_{\vec s_1 + \vec s_2 =\vec l_3} 
 \frac{b_{s_1 s_2 s_3}}{C_{s_1}} \propto A\left(\frac{\lmax}{l_{3}}\right)^{2}.
 \end{eqnarray}
 We then have that
 \begin{equation}
  \VEV{(\Delta  \mathcal{B}_1)^2}\propto \sigma_0^4 l_{\rm max}^2 A^2\sum_{\vec l_1 + \vec l_2 + \vec l_3=0} \frac{l_1^2 + l_2^2 + l_3^2}{l_2^2 l_3^4}.  \end{equation}
 We can see from this expression that it will be dominated by the terms where $l_3$ is smallest.  Therefore, taking the 
squeezed limit again we write $l_3 \ll l_2 \simeq l_1$ and find that 
\begin{equation}
\VEV{(\Delta  \mathcal{B}_1)^2} \propto \sigma_0^4 A^2\frac{l_{\rm max}^4}{l_{\rm min}^2} \propto\ln(l_{\rm max})^{-2} \label{eq:var2},
\end{equation}
where we have taken the limit $\lmax\gg l_{\rm min}$ and used the expression for $\sigma_0$ found in Eq.~(\ref{eq:fullvar}).  

The numerical calculation found in Ref.~\cite{Smith:2011rm} shows that the variance decreases slightly faster with $\VEV{(\Delta  \mathcal{B}_1)^2} \propto \ln(\lmax)^{-3}$ - non-squeezed configurations do make some contribution to the estimator which causes the variance fall off more rapidly with $\lmax$. For other forms of non-Gaussianity, the $\mathcal{O}(\fnl)$ term in the expansion of the estimator can not be written simply as in Eq.~(\ref{eq:B1}). An expression analogous to Eq.~(\ref{eq:B1Tri}), however, may still be written though with a different weighting due to the shapes of the Fourier-space triangles which dominate the $S/N$ in these cases. Since these models are not dominated by the squeezed limit, they are not afflicted by the slow scaling with $\lmax$ of the $S/N$ that occurs for the local model.

\section{Realization normalized estimator}
\label{rdn}
In order to improve the $\fnl$ estimator when $\fnl \neq 0$, we will search for a realization dependent normalization 
which removes some of the variance in those terms in the estimator which do not contribute to the signal but do 
contribute to the noise.  
 To do this we follow the approach presented in Ref.~\cite{Creminelli:2006gc} and construct a new realization-dependent normalization 
which will remove as much of the $\fnl$-dependent variance of $\mathcal{B}_1$ as possible.  In other words, we wish to define a realization-dependent normalization 
that is highly correlated with $\mathcal{B}_1$.  

We write the new realization-dependent normalization as an estimator for $\mathcal{B}_1$,  $\widehat{\mathcal{B}_1}$, so that 
we obtain a new estimator for $\fnl$, the realization-normalized estimator (RNE):
\begin{equation}
(\fnle)^{\rm N} = \frac{\fnle}{\widehat{\mathcal{B}_1}} \approx \frac{\mathcal{B}_0}{\widehat{\mathcal{B}_1}} + \fnl \frac{\mathcal{B}_1}{\widehat{\mathcal{B}_1}}.
\label{eq:newEstimator}
\end{equation}
The extent to which the new realization-dependent normalization decreases the variance of this estimator is determined by the correlation 
between $\mathcal{B}_1$ and $\widehat{\mathcal{B}_1}$; in the limit that these two terms are fully correlated the $f_{\rm NL}^2$ variance is completely removed. 

Looking at the equation for $\mathcal{B}_1$ [Eq.~(\ref{eq:B1})] the only term that cannot be immediately written in terms of observables is $a_{lm}^{\rm NL}$.  Therefore, when constructing the estimator $\widehat{\mathcal{B}_1}$ we must find a minimum variance estimator for $a_{lm}^{\rm NL}$.  To do this we 
define a weighted sum, 
\begin{equation}
\widehat{a}_{lm}^{\rm NL} = \sum_{l_{1}l_{2}}\sum_{m_{1} m_{2}} W^{m m_1 m_2}_{l l_1 l_2}a_{l_1 m_1} a_{l_2 m_2},
\end{equation}
and demand that the weight minimizes the variance, 
\begin{equation}
\frac{\delta }{\delta W^{m m_1 m_2*}_{l l_1 l_2}} \VEV{ \bigg| \widehat{a}_{lm}^{\rm NL} - a^{\rm NL}_{lm}\bigg |^2}=0.
\label{eq:minreq}
\end{equation}
Choosing the weight 
\begin{equation}
W^{m m_1 m_2}_{l l_1 l_2}=  \frac{\mathcal{G}^{m m_1 m_2}_{l l_1 l_2}}{ C_{l_1} C_{l_2}}\int r^{2} dr \alpha_{l}(r)\beta_{l_1}(r)\beta_{l_{2}}(r),
\end{equation}
satisfies Eq.~(\ref{eq:minreq}) and thus leads to a minimum variance estimator for $a^{\rm NL}_{lm}$. 
With this, the estimator for $\mathcal{B}_1$ can be written 
\begin{eqnarray}
\widehat{\mathcal{B}_1} &=&\sigma_{0}^{-2} \sum_{l_{1}\leq l_{2}\leq l_{3}}\sum_{m_{1}m_{2}m_{3}}
     \frac{\widehat{a}^{\rm NL}_{l m }a_{l_1 m_1}a_{l_2 m_2}}{C_{l} C_{l_1} C_{l_2} }B_{ll_1l_2}\wigner{l}{m}{l_{1}}{m_{1}}{l_{2}}{m_{2}}+{\rm cyclic~permutations},
     \label{eq:B1estimator}\\
     &=&\sigma_{0}^{-2} \sum_{l_{1}\leq l_{2}\leq l_{3},l_{a}l_{b}}\sum_{m_{1}m_{2}m_{3}m_{a}m_{b}}
     \frac{a_{l_1 m_1}a_{l_2 m_2}a_{l_{a}m_{a}}a_{l_{b}m_{b}}}{2C_{l} C_{l_1} C_{l_2}C_{l_{a}}C_{l_{b}} }B_{ll_1l_2}B_{ll_{a}l_{b}}\wigner{l}{m}{l_{1}}{m_{1}}{l_{2}}{m_{2}}\wigner{l}{m}{l_{a}}{m_{a}}{l_{b}}{m_{b}},
       \label{eq:B1estimator_a}
\end{eqnarray}
where we have made the approximation $a_{lm}^{\rm L}\simeq a_{lm}$, which is true to lowest order in $\fnl$.

We note that there are several other ways of arriving at the same estimator for $\mathcal{B}_1$.  In particular, if we were to instead search for a minimum variance estimator for $\Phi_{lm}$ as in 
Ref.~\cite{Komatsu:2003iq}, then we would again be lead to the estimator for $\mathcal{B}_1$ in Eq.~(\ref{eq:B1estimator_a}).  Furthermore, the same relation between 
the underlying gravitational potential and the observed multipoles 
appears when calculating the shape of the likelihood surface for the MVNH estimator as discussed in Ref.~\cite{2005PhRvD..72d3003B}.  

As we now discuss, the improved estimator given by Eqs.~(\ref{eq:newEstimator}) and (\ref{eq:B1estimator_a}) (the RNE), has two important properties.  First we show that in the case where the transfer function is just the Sachs-Wolfe transfer 
function, the RNE is the same as the estimator presented in Ref.~\cite{Creminelli:2006gc}.  Our discussion then shows how the RNE, in this limit, is able 
to remove all of the $\fnl$-dependent variance from the standard MVNH estimator.  Second, when the full transfer-function is used, the combination 
of both the Sachs-Wolfe and late-time integrated Sachs-Wolfe (ISW) effects on large angular scales causes the RNE to be less effective; it only removes 
part of the $\fnl$-dependent variance.  Using other tracers of large-scale structure, as discussed in Ref.~\cite{Mead:2010bv}, allows us to improve the performance of 
the RNE, so that it can remove \emph{most} of the $\fnl$-dependent variance.  Finally, in Appendix C we discuss how our estimator can be rewritten in terms of real-space quantities, so as to be computationally efficient. 

\section{Properties of the realization-normalized estimator}
To simplify the calculations in this section we work in the flat-sky approximation. Although the results presented will differ when compared to full-sky calculations, since we are interested in 
calculating the \emph{fractional} reduction of the $\fnl$-dependent variance (relative to the $\fnl$-dependent variance of the standard MVNH estimator), we expect the differences to be small. 

\label{varcalc}
To quantify the statistical properties of the RNE we must calculate the mean and variance of the ratios $
\mathcal{B}_0/\widehat{(\mathcal{B}_1)}$ and $\mathcal{B}_1/{\widehat{(\mathcal{B}_1)}}$. To do this we will use an approximate formula for the variance of the ratio of two stochastic variables. 
Let $x_1$ and $x_2$ be two stochastic variables with means $\mu_1$ and $\mu_2$, variances $\sigma_1^2$ and $\sigma_2^2$, and covariance $\rho$.  We wish to 
calculate the mean and variance of 
\begin{equation}
W = \frac{x_1}{x_2}.
\end{equation}
The probability density function (PDF) of $W$ may be computed analytically if $x_1$ and $x_2$ are normally distributed \cite{citeulike:4169725}.  However, the formula for the PDF
is quite complicated and it does not immediately yield analytic formulae for the mean, $\VEV{W}$, and variance $\VEV{(W-\VEV{W})^2} \equiv 
\VEV{\Delta W^2}$.  Instead, we will derive an approximate expression. 
To do this we write $x_1 = \mu_1 + \delta x_1$ and $x_2 = \mu_2 + \delta x_2$.  We then assume $\delta x_{1,2}/ \mu_{1,2} = \epsilon \ll 1$; this will be true near the peak of the normal distribution
if $\mu_{1,2} \gg \sigma_{1,2}$ (we have verified that the stochastic quantities discussed we are interested in satisfy this condition).  We can write
\begin{equation}
W \approx  \frac{\mu_1}{\mu_2}\left(1 + \delta x_1/\mu_1 - \delta x_2/\mu_2\right) + \mathcal{O}(\epsilon^2).
\end{equation}
With this we can easily compute 
\begin{eqnarray}
\VEV{W} &=& \frac{\mu_1}{\mu_2},\\
\VEV{\Delta W^2} &=& \frac{\sigma_1^2}{\mu_1^2} + \frac{\sigma_2^2}{\mu_2^2} - 2\frac{\rho}{\mu_1 \mu_2}.
\end{eqnarray}

\subsection{The mean}
We wish to show that $\left \langle \mathcal{B}_1\right  \rangle=\left \langle \widehat{\mathcal{B}_1} \right \rangle$.  To leading order in $\fnl$ in the flat-sky approximation,
\begin{eqnarray}
\VEV{\mathcal{B}_1} &=&\sigma_0^2\sum_{\vec l_1 + \vec l_2+\vec l_3=0} \delta_{\vec k + \vec k',\vec l_1} \frac{B(l_1,l_2,l_3)}{2 \Omega^2  C_{l_1} C_{l_2} } \int r^2 dr  \alpha_{l_1}(r)\VEV{a_{\vec L}^{\rm L}a_{\vec l_2}^{\rm L}\phi_{\vec k}(r) \phi_{\vec k'}(r) },\\
\VEV{\widehat{\mathcal{B}_1}} &=&\sigma_0^2\sum_{\vec l_1 + \vec l_2+\vec l_3=0} \delta_{\vec k + \vec k',\vec l_1} \frac{B(l_1,l_2,l_3)}{2 \Omega^2  C_{l_1} C_{l_2} } \int r^2 dr  \alpha_{l_1}(r) \frac{\beta_k(r) \beta_{k'}(r)}{C_k C_{k'}}\VEV{a_{\vec L}^{\rm L}a_{\vec l_2}^{\rm L}a^{\rm L}_{\vec k} a^{\rm L}_{\vec k'} }.
\end{eqnarray} 
The fact that $|\vec l_1| \geq 2$ requires that the Wick contraction which contains $\VEV{ \phi_{\vec k}(r) \phi_{\vec k'}(r)}$ 
does not contribute to the mean.  Therefore we are left with 
\begin{eqnarray}
 \VEV{a_{\vec l_2} a_{\vec L} \phi_{\vec k}(r) \phi_{\vec k'}(r)}  &=& 
\VEV{a_{\vec l_2} \phi_{\vec k}(r)} \VEV{a_{\vec L}  \phi_{\vec k'}(r)}+
 \VEV{a_{\vec L} \phi_{\vec k}(r)} \VEV{a_{\vec l_2}  \phi_{\vec k'}(r)},
  \\ &=&   \beta_{l_2}(r) \beta_{L}(r)\left[\delta_{\vec l_2, -\vec k} \delta_{\vec L, -\vec k'}+\delta_{\vec l_2, -\vec k'} \delta_{\vec L, -\vec k}   \right].
 \end{eqnarray}
Now to calculate the equivalent expression for $\widehat{\mathcal{B}}_1$.
Again, as with $\mathcal{B}_1$ only the `cross' terms survive and we have 
\begin{eqnarray}
\frac{\beta_k(r) \beta_{k'}(r)}{C_k C_{k'}} \VEV{a_{\vec l_2} a_{\vec L} a_{\vec k} a_{\vec k'}} 
=  
 \beta_{l_2}(r) \beta_{L}(r) \left[\delta_{\vec l_2, -\vec k} \delta_{\vec L, -\vec k'}+\delta_{\vec l_2, -\vec k'} \delta_{\vec L, -\vec k}   \right],
\end{eqnarray}
so we have that $\VEV{\mathcal{B}_1} = \VEV{\widehat{\mathcal{B}}_1}$. 
With this we can conclude that the RNE is unbiased. 

\subsection{The variance}

The variance of the first term
 can be approximated by 
\begin{eqnarray}
\VEV{\left(\Delta \frac{(\mathcal{B}_0)}{\widehat{(\mathcal{B}_1)}}\right)^2}&\approx&
\VEV{ (\mathcal{B}_0)^2}+ \VEV{ (\mathcal{B}_0)^2\Delta \widehat{(\mathcal{B}_1)} \Delta  \widehat{(\mathcal{B}_1)}} \\
&\approx&
\VEV{ (\mathcal{B}_0)(\mathcal{B}_0)}\left(1+ \VEV{\Delta \widehat{(\mathcal{B}_1)} \Delta  \widehat{(\mathcal{B}_1)}}\right) ,\\
&=&\sigma_0^{2}\left(1+ \VEV{\Delta \widehat{(\mathcal{B}_1)} \Delta  \widehat{(\mathcal{B}_1)}}\right).
\end{eqnarray}
The variance of the second term can be written as  
\begin{eqnarray}
\VEV{\left(\Delta \frac{(\mathcal{B}_1)}{\widehat{(\mathcal{B}_1)}}\right)^2}\approx \VEV{\Delta (\mathcal{B}_1)\Delta(\mathcal{B}_1)}+ \VEV{\Delta \widehat{(\mathcal{B}_1)}\Delta \widehat{(\mathcal{B}_1)}}- 
2\VEV{\Delta (\mathcal{B}_1)\Delta \widehat{(\mathcal{B}_1)}}.
\end{eqnarray}
A tedious yet straight-forward calculation shows that the variance of the RNE is composed of four terms (we show the details of the calculation in Appendix A).  Of those
four terms one dominates in the $\lmax \gg 1$ limit so that we have 
\begin{eqnarray}
\VEV{\Delta (\mathcal{B}_1)\Delta(\mathcal{B}_1)} &=& 8\sigma_{0}^{4} \sum_{\{\vec l\},\{\vec k\}} \frac{B(l_1,l_2,l_3) B(k_1,k_2,k_3) }{C_{l_1} C_{l_2} C_{k_1} C_{k_2} C_{k_3}}\delta_{\vec l_3, \vec k_3}  \\
&\times& \int r^2 dr (r')^2 dr' \alpha_{l_1}(r) \alpha_{k_1}(r')\beta_{k_2}(r') \beta_{l_2}(r) \chi_{l_3}(r,r') ,\nonumber \\
\VEV{\Delta \widehat{(\mathcal{B}_1)}\Delta \widehat{(\mathcal{B}_1)}} &=&
 8 \sigma_{0}^{4}\sum_{\{\vec l\},\{\vec k\}} \frac{B(l_1,l_2,l_3) B(k_1,k_2,k_3) }{C_{l_1} C_{l_2} C_{k_1} C_{k_2} C_{k_3}}\delta_{\vec l_3, \vec k_3}  \\
&\times& \int r^2 dr (r')^2 dr' \alpha_{l_1}(r) \alpha_{k_1}(r')\beta_{k_2}(r') \beta_{l_2}(r) \frac{\beta_{l_3}(r) \beta_{l_3}(r')}{C_{l_3}},\nonumber \\
\VEV{\Delta (\mathcal{B}_1)\Delta \widehat{(\mathcal{B}_1)}}&=& \VEV{\Delta (\mathcal{B}_1)\Delta(\mathcal{B}_1)}.
\end{eqnarray}
The variance is given by
\begin{eqnarray}
\left \langle \left( \Delta \frac{\mathcal{B}_{1}}{\widehat{\mathcal{B}}_{1}}\right)^{2}\right \rangle &=& 8 \sigma_{0}^{4}\sum_{\{\vec l\},\{\vec k\}} \frac{B(l_1,l_2,l_3) B(k_1,k_2,k_3) }{2C_{l_1}  C_{l_2} C_{k_1} C_{k_2} C_{k_3}}\delta_{\vec l_3, \vec k_3} \label{eq:var3} \nonumber \\
&\times& \int r^2 dr (r')^2 dr' \alpha_{l_1}(r) \alpha_{k_1}(r') \beta_{l_2}(r) \beta_{k_2}(r')\mathcal{D}_{l_{3}}(r,r'),\\
\mathcal{D}_{l}(r,r')&\equiv&\left[\chi_{l}(r,r') -\frac{\beta_{l}(r)\beta_{l}(r')}{C_{l}}\right].\label{eq:var4}
\end{eqnarray}

We are interested in calculating the \emph{fractional reduction} of the $\fnl$-dependent variance in the RNE, $(\fnle)^{\rm N}$, relative to the $\fnl$-dependent variance in the standard MVNH estimator, $\fnle$.  To quantify this we will define 
\begin{equation}
\mathcal{R} \equiv {\left \langle \left( \Delta \frac{\mathcal{B}_{1}}{\widehat{\mathcal{B}}_{1}}\right)^{2}\right \rangle}\bigg /{\left \langle  \Delta\mathcal{B}_{1}^2\right \rangle  }.
\end{equation}
If $\mathcal{R} = 0$ then all of the $\fnl$-dependent variance has been removed; for $0<\mathcal{R}<1$ then there is a residual $\fnl$-dependent variance which the RNE 
does not remove. 

 \begin{figure}[ht]
\begin{center}
\resizebox{!}{7cm}{\includegraphics{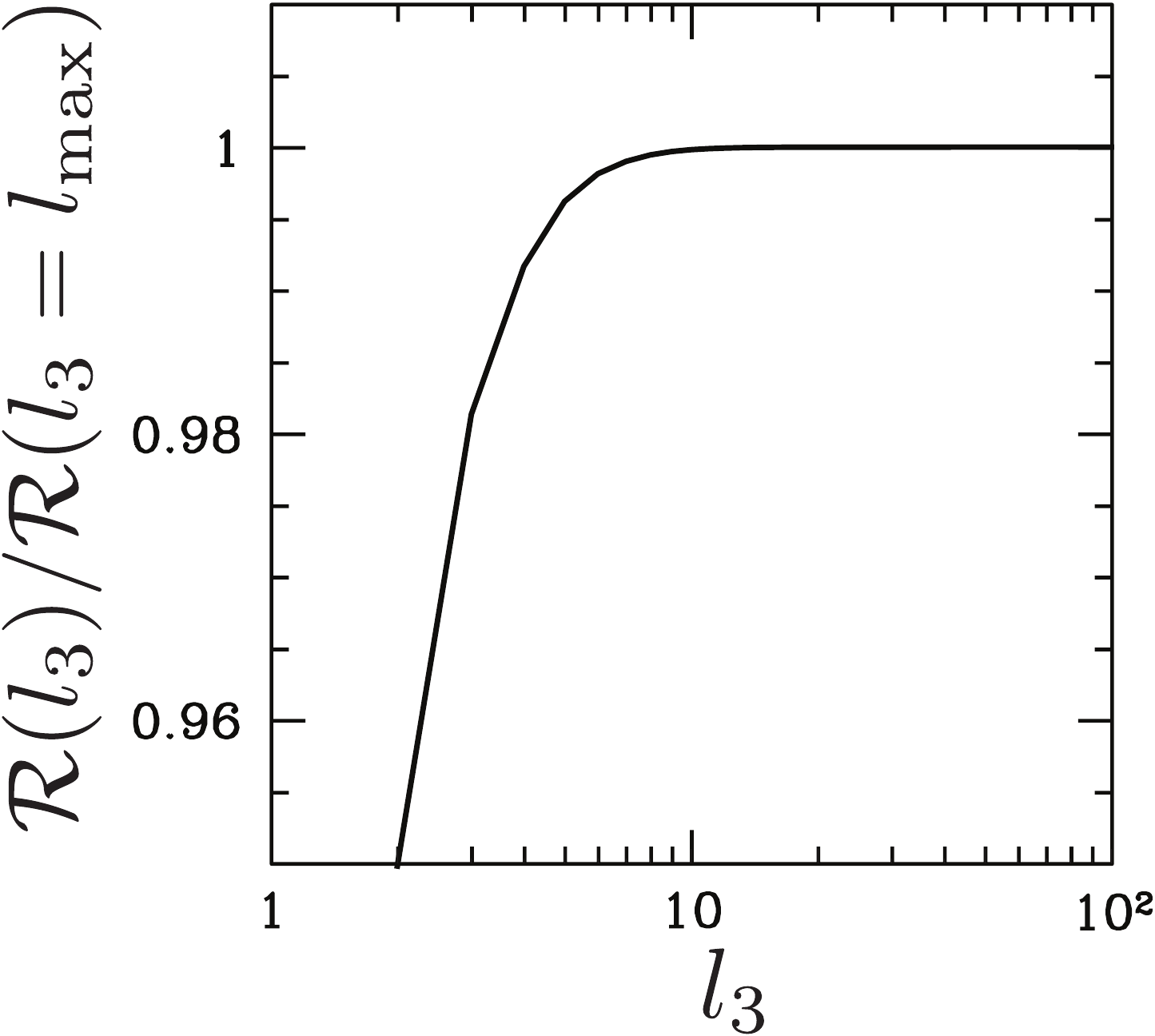}}
\caption{The ratio of the fractional reduction of the $\fnl$-dependent variance in $(\fnle)^{\rm N}$ as a function of the value of the $l_3$ multipole (which is the large scale in the squeezed limit).  This figure 
was produced with $\lmax = 100$ (we have checked that other choices for $\lmax$ reproduce curve).  From this figure we can see that $\sim 95\%$ of the fractional reduction, $\mathcal{R}$, is contained in the quadrupole $l_3 = 2$ justifying 
our approximation in Eq.~(\ref{eq:approx1}). }
\label{fig:frac}
\end{center}
\end{figure}
Looking at Eq.~(\ref{eq:var3}) we can see that the extent to which we reduce the variance of the RNE depends on the function $D_l(r,r')$.  In
the squeezed limit ($|\vec l_3| \sim l_{\rm min} \ll |\vec l_1| \simeq |\vec l_2|$ and $|\vec k_3| \sim l_{\rm min} \ll |\vec k_1| \simeq |\vec k_2|$), 
we have  
\begin{eqnarray}
\left \langle \left( \Delta \frac{\mathcal{B}_{1}}{\widehat{\mathcal{B}}_{1}}\right)^{2}\right \rangle &\propto&\sum_{l_3}\mathcal{D}_{l_3}(r_*,r_*) \sum_{\{\vec l\},\{\vec k\}} \frac{B(l_1,l_2,l_3) B(k_1,k_2,l_3) }{C_{l_1}  C_{l_2} C_{l_3}C_{k_1} C_{k_2} }  \beta_{l_2}(r_*) \beta_{k_2}(r_*),
\end{eqnarray}
where we have made the substitution $r^2 \alpha_l(r) \propto \delta(r-r_*)$ for $l \gg l_{\rm min}$-- i.e., to a very good approximation, all small-scale power is sourced by physics that occurs at the surface of last scattering.  This allows us to write the fractional reduction of the $\fnl$-dependent variance as 
\begin{equation}
\mathcal{R}
\approx \frac{\mathcal{D}_2(r_*,r_*)}{\chi_{2}(r_*,r_*)},
\label{eq:approx1}
\end{equation}
where we have fixed $l_3 = l_{\rm min} = 2$ since that value contains upwards of 95\% of the variance (see Fig.~\ref{fig:frac}).  We have verified this approximation by 
calculating the full sum in Eq.~(\ref{eq:var3}) and found it to be accurate to a few percent. The function $\chi_{l}(r,r^{\prime})$ is calculated using $P(k)=A k^{n_{\rm s}-4}$, where we use $n_{s}=1$ to simplify the computation of the integral over $k$ (the dependence on the actual value of $n_s$ is negligible). 

\subsection{The $\fnl$-dependent variance in the Sachs-Wolfe limit}
Before calculating the improvement in the $\fnl$-dependent variance 
in the case of a $\Lambda$CDM universe, let us first analyze these expressions in the Sachs-Wolfe limit. In this limit $\alpha_l(r) \propto \delta(r-r_*)$.  From Eq.~(\ref{eq:var4}) it is clear that the $\fnl$-dependent variance will vanish (i.e., $\mathcal{R} = 0$) since it depends on 
\begin{equation}
\mathcal{D}_l(r_*,r_*) = \chi_l(r_*,r_*) - \beta_l^{2}(r_*)/C_l = 0.
\end{equation}
This result was found in Refs.~\cite{Creminelli:2006gc,Liguori:2007sj,Smith:2011if}.  

We can understand this result in a slightly different way which will highlight how the Sachs-Wolfe limit is unique.  In the Sachs-Wolfe limit the estimator $\widehat{a}_{lm}^{\rm NL}$ is \emph{fully correlated} (up to order $A f_{\rm NL}^2$) with $a_{lm}^{\rm NL}$ since 
\begin{eqnarray}
\widehat{a}_{lm}^{\rm NL} &=&-3 \sum_{l_{1}l_{2}}\sum_{m_{1}m_{2}} (-1)^m\mathcal{G}^{-m m_1 m_2}_{l l_1 l_2}a_{l_1 m_1} a_{l_2 m_2} ,\\
&=&-\frac{1}{3} \sum_{l_{1}l_{2}}\sum_{m_{1} m_{2}}(-1)^m \mathcal{G}^{-m m_1 m_2}_{l l_1 l_2}\phi_{l_1 m_1}(r_*) \phi_{l_2 m_2}(r_*) + \mathcal{O}(A f_{\rm NL}^2)\\
a^{\rm NL}_{lm}&=& \sum_{l_{1}l_{2}}\sum_{m_{1}m_{2}}(-1)^m\mathcal{G}^{-m m_1 m_2}_{l l_1 l_2}\int r^{2} dr \alpha_{l}(r)\phi_{l_{1}m_{1}}(r)\phi_{l_{2}m_{2}}(r) \\
&=& -\frac{1}{3} \sum_{l_{1}l_{2}}\sum_{m_{1}m_{2}}(-1)^m \mathcal{G}^{-m m_1 m_2}_{l l_1 l_2}\phi_{l_1 m_1}(r_*) \phi_{l_2 m_2}(r_*).
\end{eqnarray}

In a $\Lambda$CDM universe, however, the large-scale anisotropies receive power from \emph{both} the Sachs-Wolfe as well as the 
late-time integrated Sachs-Wolfe (ISW) effects. This additional contribution to the large-scale power degrades the correlation between $\widehat{a}_{lm}^{\rm NL}$ and  $a_{lm}^{\rm NL}$ thus leaving some residual $\fnl$-dependent variance.   
 
\subsection{The $\fnl$-dependent variance for the full $\Lambda$CDM transfer function}
 We are now in a position to calculate the reduction of the $\fnl$-dependent variance using the realization-dependent normalization in a $\Lambda$CDM universe. As we already discussed, the value of the ratio $\mathcal{D}_l(r_*,r_*)/\chi_l(r_*,r_*)$ at the quadrupole ($l=2$) gives a good estimate for the fractional reduction of the $\fnl$-dependent variance, $\mathcal{R}$.  

\begin{figure}[ht]
\begin{center}
\resizebox{!}{7cm}{\includegraphics{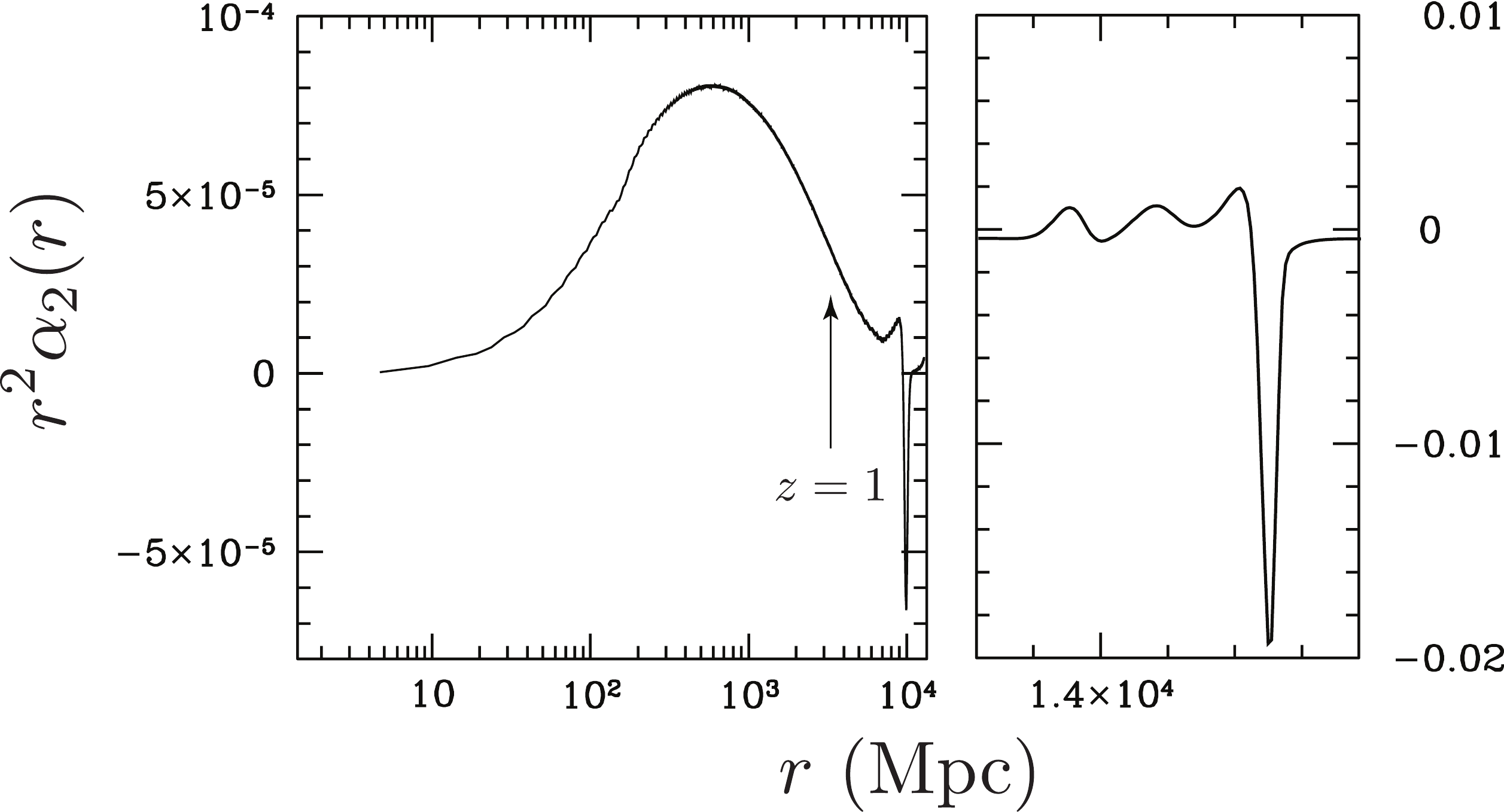}}
\caption{The transfer function, $r^2 \alpha_l(r)$, for the quadrupole ($l=2$) as a function of conformal distance in Mpc.  The left-hand panel 
shows the effects of late-time ISW.  The dip around $r \approx 10^4$ Mpc corresponds to reionization at $z = 11$.  The right-hand panel highlights the evolution of $r^2 \alpha_l(r)$ around decoupling.  Note that the 
scale is not the same in both panels.}
\label{fig:alphaLCDM}
\end{center}
\end{figure}

In Fig.~\ref{fig:alphaLCDM} we show how the quadrupole filter $\alpha_2(r)$ [see Eq.~(\ref{eq:alpha})] depends on the line-of-sight distance $r$ in a $\Lambda$CDM 
universe.  First note that there is a relatively large rise below $r \leq 10^3$ Mpc, corresponding to the late-time ISW effect.  Furthermore at $r \simeq 2 \times 10^4$ Mpc, the filter shows a sharp dip corresponding to the 
contribution from the surface of last scattering.  Therefore in a $\Lambda$CDM universe on large scales the observed multipoles $a_{lm}$ have contributions from two separate epochs: the ISW effect at late-times and the Sachs-Wolfe effect around the surface of last scattering.  This two-epoch contribution is central to understanding how the RNE works when applied to a $\Lambda$DCM universe.

 \begin{figure}[ht]
\begin{center}
\resizebox{!}{7cm}{\includegraphics{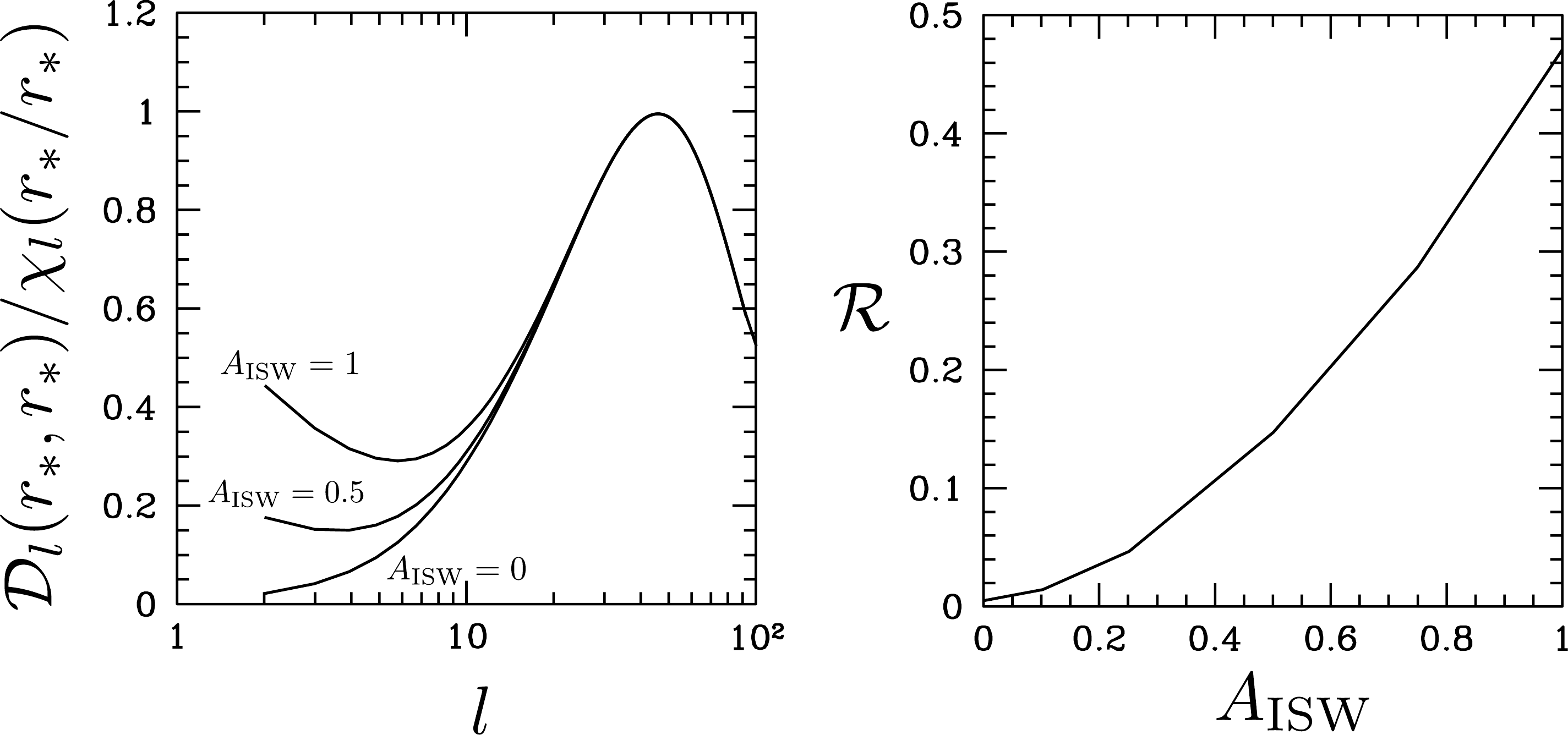}}
\caption{\emph{Left}: The realization-dependent normalization reduces the $\fnl$-dependent variance to varying extents depending on the level of the late-time ISW effect.  From top to bottom we plot the function $D_l(r_*,r_*)/\chi_l(r_*,r_*)$ with varying levels of late-time ISW: 1, 0.5, 0.  The overall reduction of the $\fnl$-dependent variance roughly corresponds to the value of $\mathcal{D}_2(r_*,r_*)/\chi_2(r_*,r_*)$. \emph{Right}: The fractional improvement in the $\fnl$-dependent variance, $\mathcal{R}$, as a function of the amplitude of the late-time ISW.  When the late-time ISW is completely removed (i.e., $A_{\rm ISW} = 0$) the 
$\fnl$-dependent variance is nearly completely removed; for an unmodified ISW level the $\fnl$-dependent variance is reduced by a factor of $\sim0.5$.}
\label{fig:Dl}
\end{center}
\end{figure}

To investigate how the late-time ISW effect impacts the improvement of the RNE, we modified the publicly available Boltzmann code CAMB \cite{lewis2000} to include a 
new parameter, $A_{\rm ISW}$, which controls the amplitude of the late-time ISW effect in the line-of-sight integral \cite{Seljak:1996is}.  When $A_{\rm ISW}=1$ it takes on 
its standard value in the calculation of the $C_l$s; when it is zero the late-time ISW effect is absent. In Fig.~\ref{fig:Dl}
we show $\mathcal{D}_l(r_*,r_*)/\chi_l(r_*,r_*)$ as a  function of multipole, $l$, with varying late-time ISW  amplitude, $A_{\rm ISW}$.
On the right hand side of the figure, we can see $\mathcal{R} \simeq \mathcal{D}_2(r_*,r_*)/\chi_2(r_*,r_*)$ falls off sharply as $A_{\rm ISW}$ decreases. 
Qualitatively this result can be understood by noting that with both the Sachs-Wolfe and late-time ISW effects contributing to the large-scale anisotropies our estimator  $\widehat{a}_{lm}^{\rm NL}$ becomes less correlated with the actual $a_{lm}^{\rm NL}$ leading to a larger value for the fractional reduction $\mathcal{R}$.

In order to better understand how the late-time ISW effect limits our ability to remove the $\fnl$-dependent variance quantitatively we can approximate the anisotropies on large-scales by 
\cite{1995PhDT..........H}
\begin{equation}
a_{lm} \simeq - \frac{1}{3} \Phi_{lm}(r_*) +A_{\rm ISW} \Phi_{lm}(r_{\rm ISW}), 
\label{eq:ISWalm}
\end{equation}
where $A_{\rm ISW}$ controls the contribution of the ISW effect to the temperature multipoles, and $r_{\rm ISW}$ is the conformal distance from the observer to the peak of the ISW visibility function. The ISW contribution extends from $\sim$ 500 Mpc to $4000$ Mpc (where the present time is $0$ Mpc). 
Eq.~(\ref{eq:ISWalm}) explicitly shows how the contribution of the late-time ISW effect to the observed multipole moments $a_{lm}$ limits our ability to reconstruct $a_{lm}^{\rm NL}$.
Since the correlation between the Sachs-Wolfe and ISW terms is at most 10\% we neglect it and we have 
\begin{equation}
\mathcal{R} \simeq \frac{\mathcal{D}_2(r_*,r_*)}{\chi_2(r_*,r_*)} \simeq 1 - \frac{C_l^{\rm SW}}{C_l^{\rm SW}+ A_{\rm ISW}^2C_2^{\rm ISW}},
\end{equation} 
where $C_l^{\rm SW}$ and $C_l^{\rm ISW}$  are the Sachs-Wolfe and the late-time ISW contribution to the power spectrum, respectively.  
As shown on the right-hand panel in Fig.~\ref{fig:Dl}, in a $\Lambda$CDM universe with all of the late-time ISW effect included, the $\fnl$-dependent variance can only be reduced by a factor of 0.5 using the realization-dependent normalization defined in Eqs.~(\ref{eq:newEstimator}) and (\ref{eq:B1estimator}).  When the late-time ISW contribution is completely removed, the $\fnl$-dependent variance is reduced by a factor of 20 ($\mathcal{R}= 0.05$). This raises the question of whether or not probes of large-scale structure closer to the present epoch can be used to `remove' the late-time ISW contribution to the CMB anisotropies and further reduce the $\fnl$-dependent variance. Next we discuss how we may use such a tracer of the ISW effect to further reduce the $f_{\rm NL}$-dependent variance of the RNE.

\section{ISW subtraction with foreground tracers}
\label{iswsub}

We have identified the late-time ISW effect as the cause of the residual variance scaling as $f_{\rm NL}^{2}$. If the late-time ISW component of a temperature map could be estimated, and then cleaned out from the data, we expect that the corresponding generalization of the RNE would then be nearly free of the $f_{\rm NL}$-dependent variance. 

Since the bulk of the late-time ISW effect in a $\Lambda$CDM cosmology comes from redshift $z\sim 1$, a measurement of the large-scale gravitational potential or density field around this redshift should yield information that we can use to clean our map of the ISW effect. The use of an ISW-cleaned map to improve the sensitivity of the CMB to primordial non-Gaussianity was first proposed in Ref.~\cite{Mead:2010bv}. In that work, only the zeroth order variance $\left \langle(\Delta \mathcal{B}_{0})^{2}\right\rangle$ was computed, and as a result the improvement in the $S/N$ obtained by using ISW-cleaned maps was marginal. In contrast, here we will see that using a large-scale structure tracer has the potential to reduce the $\fnl$-dependent variance by up to $\sim$ 90\%.

For an arbitrary large-scale tracer $t_{lm}$, the ISW-cleaned CMB temperature anisotropy is
\begin{equation}
a_{lm}^{\rm c}=a_{lm}-\frac{\left \langle a_{lm}^{\rm ISW} t_{lm}^{*}\right\rangle}{\left \langle t_{lm}~t_{lm}^{*}\right \rangle}t_{lm},
\end{equation}
where $a_{lm}^{\rm ISW}$ is the portion of the total multipoles, $a_{lm}$, due to the late-time ISW effect, $t_{lm}$ is the multipole associated with the large-scale tracer field, a superscript $c$ indicates a quantity that has been `cleaned' of the late-time ISW effect, and a superscript $t$ indicates a quantity associated with the tracer field.  We note that if we use the lensing potential for CMB weak lensing as our tracer field then $t_{lm}$ is obtained from some higher-order cumulant of the data.  This then implies that the tracer's power spectrum must also take into account an additional noise term, as we discuss in more detail below. 

The RNE takes the same form as in Eq.~(\ref{eq:newEstimator})
but written in terms of ISW-cleaned quantities:
\begin{eqnarray}
C_{l}^{\rm c}&\equiv& C_{l}-2\frac{C_{l}^{\rm T,t}C_{l}^{\rm ISW, t}}{C_{l}^{\rm tt}}+\frac{\left(C_{l}^{\rm ISW,t}\right)^{2}}{C_{l}^{\rm tt}},\\
\alpha_l^{\rm c}(r) &\equiv& \alpha_l(r) - \alpha_l^{\rm t}(r) ,\\
\beta_l^{\rm c}(r) &\equiv& \beta_l(r) - \beta_l^{\rm t}(r),
\end{eqnarray}
where $\alpha_l^{\rm t}(r)$ and $\beta_l^{\rm t}(r)$ are defined as in Eqs.~(\ref{eq:alpha}) and (\ref{eq:beta}) but with in terms of a transfer function corresponding 
to the power spectrum of the tracer field. 
In order to evaluate the fractional reduction of the $f_{nl}$-dependent variance with an ISW cleaned CMB map, we evaluate Eq.~(\ref{eq:var3}) making the identification $\alpha_{l}(r)\to \alpha_{l}^{\rm c}(r)$, $\beta_{l}(r)\to \beta_{l}^{\rm c}(r)$, and $C_{l}\to C_{l}^{c}$. Now that we have expressions for the fractional reduction of the $\fnl$-dependent variance of an estimator $\widehat{f_{\rm NL}}^{\rm c}$ built from ISW-cleaned maps, we consider two specific tracers of foreground structure: weak lensing of the CMB and galaxy surveys. 

\subsection{Reconstructed lensing potential}
 \begin{figure}[ht]
\begin{center}
\resizebox{!}{5.5cm}{\includegraphics{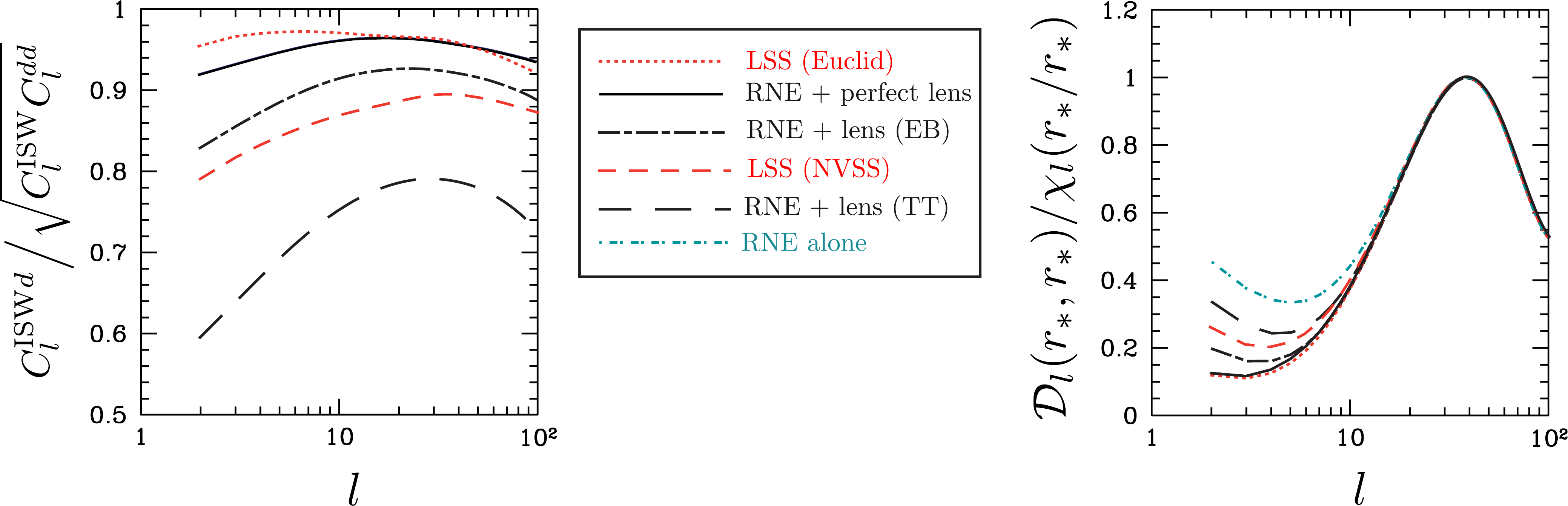}}
\caption{\emph{Left}: The three black curves show the correlation between the lensing potential and the late-time ISW effect in the CMB temperature: the solid curve shows the correlation without taking into account lensing reconstruction noise; the dot-dashed curve shows the correlation between the late-time ISW effect and the reconstructed lensing potential using EB correlations, including reconstruction noise; analogously the long-dashed curve shows the same correlation, but using TT correlations to reconstruct the lensing potential. The two red curves show the correlation between the forecasted Euclid/WFIRST (dotted) or NVSS (medium dashed) density fields, and the late-time ISW effect. \emph{Right}: The fractional reduction in the quantity $\mathcal{D}_l$ for the uncleaned CMB maps (cyan, medium-short dashed) and cleaned maps with line-types corresponding to the curves in the left panel.}
\label{fig:corr}
\end{center}
\end{figure}

Weak lensing deflects the trajectories of CMB photons, remapping the temperature and polarization fields. The projected potential $\Psi(\hat{n})$ determines the trajectories of lensed CMB photons. A reconstruction of the projected potential field from measurements of the 
CMB would allow a separate probe of large-scale structure up to redshifts of $\sim$ a few. 

Referring to the solid curves in the left-hand panel of Fig.~\ref{fig:corr} we show the correlation coefficient between the lensing potential and the late-time ISW effect, as well as the correlation coefficient between the deflection and full temperature fields. We see that the multipole moments of the deflection field are very strongly correlated with the late-time ISW effect. The fractional reduction of variance obtained with the realization-dependent normalization for the perfectly (via lensing) cleaned maps is controlled by the ratio $\mathcal{D}^{\Psi}_{l}(r_*,r_*)/\chi_{l}(r_*,r_*)$ is shown in the right-hand panel of Fig.~\ref{fig:corr}. We find that the realization-dependent normalization removes $\sim$ 90\% of the $f_{\rm NL}^{2}$ variance if the map is `cleaned' of the late-time ISW effect using a perfect reconstruction of the CMB deflection potential. 

Of course, even with a idealized noiseless CMB map, the deflection potential can not be perfectly reconstructed. The lensing estimator relies on off-diagonal correlations of temperature multipoles, and in any given realization of the power spectrum, there will be some overlap between the lensing estimator and chance correlations that arise from cosmic variance. This leads to the reconstruction noise variance $N_{l}$ \cite{Mead:2010bv,Hu:2000ee}. 

On the large scales relevant for ISW subtraction, the bulk of the $S/N$ for lensing construction comes from very high $l$, and so instrument noise can be neglected in our estimates.  We make the replacement $C_{l}^{\Psi \Psi}\to C_{l}^{\Psi \Psi}+N_{l}$ above to assess how much a realistic reconstruction of $\Psi$ (and subsequent cleaning of the CMB temperature map) could improve the performance of the RNE.  We assume a reference experiment with temperature and polarization noise $\Delta T = \Delta P/\sqrt{2} = 1\ \mu{\rm K}\ {\rm arcmin}$, and an angular resolution of $\sigma = 4'$, as discussed in detail in Ref.~\cite{2003PhRvD..67h3002O}.  We note that measurements using Planck will not be sensitive enough to realistically use the reconstructed lensing potential in order to remove the late-time ISW effect. From the right-hand panel of Fig.~\ref{fig:corr}, we see that if the temperature field is used as to reconstruct the deflection field, the RNE can remove $\sim 70 \%$ of the variance proportional to $f_{\rm NL}^{2}$, whereas if the polarization field (through the EB correlation) is used, the RNE can remove $\sim 80 \%$ of the excess variance\footnote{Of course in a more complete analysis, we could include other correlation functions (EE, TE, EE, and BB) in the reconstruction of the deflection field. We choose to focus on TT (since TT-based detection of CMB weak lensing have already been made) and EB (because it vanishes in the null hypothesis of no lensing, and thus provides a strong probe of the deflection field). These other correlation functions could themselves be used to look for non-Gaussianity-- indeed, Refs. \cite{Babich:2004yc} and \cite{2008ApJ...678..578Y} introduce the MVNH that includes all of them. In this work, we restrict ourself to the simple case discussed already and leave a more complete analysis for future work.}.

\subsection{Galaxy survey}
Alternatively, a galaxy survey with peak redshift near $z\sim 1$  also probes the potential field at the epochs when the late-time ISW effect is imprinted on the CMB. For a galaxy survey with selection function $w(z)$, the transfer-function is given by
\begin{equation}
S^g_l(k) \equiv \int_0^{r_*} dr b_k(r) S^m_{k}(r) w[z(r)] j_{l}(kr)
\end{equation}
where $b_k(r)$ is the bias, $z(r)$ is the redshift as a function of conformal distance, and $S_{k}^{m}(r)$ is the time and scale-dependent function which maps the primordial potential to the evolved matter density. We have taken \cite{2008PhRvD..77l3520G,2010MNRAS.406....2F}
\begin{equation}
w(z)= CH(z)\left( \frac{z^m}{z_0^{m+1}}\right) \exp\left[-\left(\frac{z}{z_0}\right)^{\beta}\right], 
\end{equation}where for the NVSS survey we take $z_{0}=0.79$, $\beta=1$, and $m=1.18$ as in Ref.~\cite{Ho:2008bz}. On the other hand, for a futuristic space-based all-sky dedicated dark energy, such as that described in the Joint Dark Energy Mission (WFIRST) \cite{Joudaki:2011nw} and Euclid mission concept \cite{Schaefer:2009yc}, or the Large Synoptic Survey Telescope \cite{Joudaki:2011nw} (LSST) we take $z_{0}=0.5$, $\beta=1$, $m=2$ \cite{Joudaki:2011nw}. The presence of the Hubble parameter converts the selection function from galaxies per unit redshift, to galaxies per unit conformal time.  We do not need to specify the normalization constant $C$, as it divides out of all quantities of interest.

We approximate the bias as constant in scale as well as in redshift \cite{2008PhRvD..77l3520G,2010MNRAS.406....2F} so that it factors out of the computation completely\footnote{In reality, local-type non-Gaussianity would also induce scale-dependent bias \cite{Dalal:2007cu}, and thus corrections to the variance of the cleaned map. This effect, however, would be higher order in $f_{\rm NL}$, so we may neglect it without loss of generality.}. Referring to the dashed-lines in Fig.~\ref{fig:corr}, we show the correlation coefficient between large-scale structure (for both NVSS and Euclid/WFIRST parameters) and the late-time ISW effect, as well as the correlation coefficient between large-scale structure and the full temperature field. We see that the multipole moments of the large-scale density field are very strongly correlated with the late-time ISW effect. 

The fractional reduction in the variance obtained with the RNE for the cleaned maps is shown in the right-hand panel of Fig.~\ref{fig:corr}. We find that the realization-dependent normalization removes $\sim 75$\% of the $f_{\rm NL}^{2}$ variance if the ISW effect is removed by cross-correlating a CMB map with a large-scale structure survey for the NVSS survey parameters, and $\sim 90\%$ of the $f_{\rm NL}^{2}$ variance if WFIRST survey parameters are assumed. In Fig.~\ref{fig:SN}, we see how the $S/N$ for $\fnl$ is improved if prior to the application of the realization-dependent normalization, the late-time ISW effect is removed from maps using two different secondary tracers.  This figure was produced assuming the variance of the standard MVNH estimator given in Eq.~(\ref{eq:fullvar}) and the reduction in the $\fnl$-dependent variance, $\mathcal{R}$, using various large-scale structure tracers to 
`clean' the CMB maps of the late-time ISW effect, 
\begin{equation}
\frac{S}{N} = \frac{\fnl}{\sqrt{\frac{1}{72 A f_{\rm sky} l_{\rm max}^2 \ln(l_{\rm max})}+\mathcal{R} \frac{ f_{\rm NL}^2}{ 2 \ln^3(l_{\rm max})}}}.
\end{equation}
Although the scaling with $l_{\rm max}$ and $\fnl$ given in the above equation was calculated in the flat-sky Sachs-Wolfe limit \cite{Creminelli:2006gc, Smith:2011rm,yd}, it
 has been shown to reproduce the $\lmax$ scaling calculated on the full sky and with the full transfer-function \cite{Babich:2004yc,Liguori:2007sj}.  However, we note that the exact numerical factors may be different for the full sky and full transfer-function case. It is thus possible that the $\fnl$-dependent variance may be even more important at lower $f_{\rm NL}$ than is indicated by this expression.  As future data is used to determine the level of non-Gaussianity in the CMB, the statistics of the standard null-hypothesis estimator should be checked, especially if the data starts to indicate that $\fnl \neq 0$.  

We see from this figure that if primordial non-Gaussianity is of local type and $\fnl\gtrsim 5$, a significant improvement in the $S/N$ for $\fnl$ may be obtained by using cleaned maps.  Of course in a real galaxy survey, there are additional complications due to incomplete sky coverage, photometric redshift errors, and shot-noise due to a finite number of galaxies in the survey volume. Here we neglect these important real-world effects to highlight the fact that a measurement of the ISW effect can help reduce the $f_{\rm NL}^{2}$ variance in the RNE.  An increase in the $S/N$ in an estimate for $\fnl$ would not only lead to a more precise determination of the level of non-Gaussianity in the CMB, but would also lead to a
more precise estimation of additional parameters (such as a possible scale-dependence) associated with primodial non-Gaussianity. 

\section{Conclusions}

We have investigated the origin of the $\fnl$-dependent variance when applying the standard MVNH $\fnl$ estimator to CMB maps with appreciable non-Gaussianity.  We found 
that this variance is due to terms that appear in the estimator which do not contribute to the signal but which do contribute to the noise.  

Previous work in Ref.~\cite{Elsner:2010hb} has shown that a Bayesian analysis has the potential to provide an estimate of $\fnl$ from the CMB which does not show an $\fnl$-dependent increase in the variance when applied to maps with appreciable non-Gaussianity.  That approach, however, is computationally expensive and quite inefficient, taking 150,000 CPU hours to compute the estimator on a simulated non-Gaussian CMB map.  It is therefore desirable to find a computationally simple and efficient method to estimate $\fnl$ from the CMB which remains optimal even when applied to maps with appreciable non-Gaussianity. 
\begin{figure}[h!]
\begin{center}
\resizebox{!}{7cm}{\includegraphics{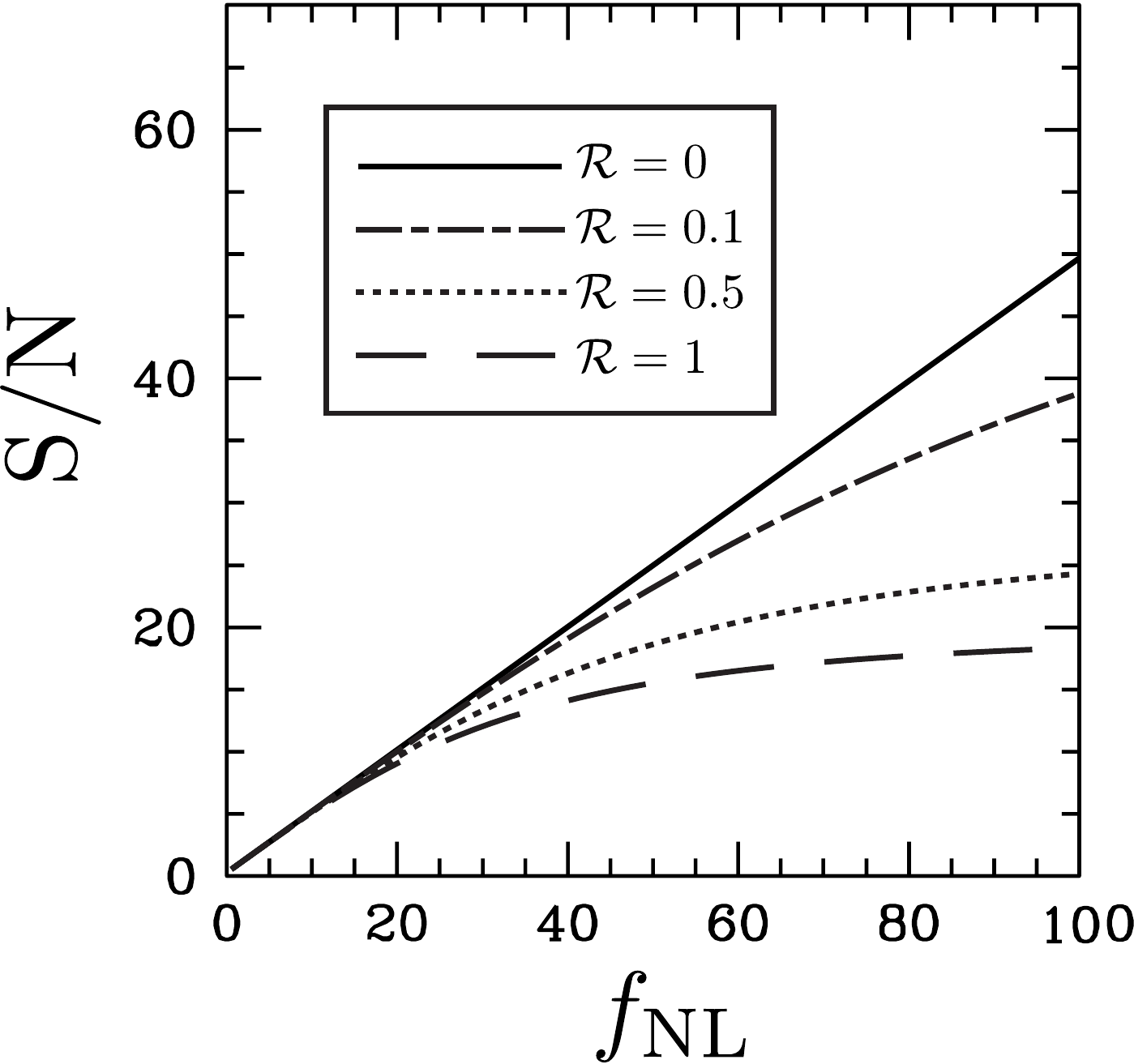}}
\caption{The signal to noise ($S/N$) in an experiment with $\lmax = 2500$ and $f_{\rm sky} = 0.8$ (corresponding to Planck) as a function of $\fnl$ under the flat-sky and Sachs-Wolfe approximations.  The top (solid) curve shows the $S/N$ for the standard MVNH estimator without taking into account the $\fnl$-dependent variance.  The curve second from the top (dot-dashed) shows the $S/N$ for the RNE after a tracer field has been used to remove most of the late-time ISW effect, leading to a reduction of the $\fnl$-dependent variance by a factor of $\mathcal{R} = 0.1$. The curve third from the top (dotted) shows the $S/N$ for the RNE using only CMB data with $\mathcal{R} = 0.5$.  The bottom curve (long-dashed) shows the $S/N$ for the standard MVNH estimator with the (full) $\fnl$-dependent variance with $\mathcal{R} = 1$.  As this figure shows, the cleaned maps can increase the $S/N$ by several standard deviations.  }
\label{fig:SN}
\end{center}
\end{figure}

We have found that a new realization-dependent estimator (RNE) can be constructed which is computationally efficient (utilizing the scaling properties of fast-Fourier transforms) and 
reduces the $\fnl$-dependent variance by a factor of $\sim 2$.  Previous studies have shown this same realization-dependent normalization can completely remove the $\fnl$-dependent variance 
in the Sachs-Wolfe limit. When the full transfer-function is used, however, this limit is a poor approximation to the CMB power-spectrum, especially at 
large-scales where the late-time ISW effect (due to late-time acceleration) contributes about half of the power.  We have artificially reduced the level of the late-time ISW 
and found that when it is completely removed our new estimator has negligible $\fnl$-dependent variance even with the full transfer function.  
This implies that by using a tracer which effectively removes the late-time ISW contribution to the CMB map we can use the RNE to 
further reduce the $\fnl$-dependence. 

We considered two tracers of large-scale structure: the deflection field (which generates lensing in the CMB) and a large-scale structure survey (considering survey parameters comparable to those of NVSS and Euclid/WFIRST) with a mean redshift of $z\sim 1$.  Both tracers are highly correlated with the late-time ISW.  We find that by using the deflection field as a tracer, we can reduce the $\fnl$-dependent variance by a factor of $\sim 0.1$ using a futuristic CMB experiment (the reconstruction of the deflection 
potential from Planck is not accurate enough to be useful). If the large-scale structure measured by NVSS is used as a tracer, the variance could be reduced by a factor of $\sim 0.25$, while a next-generation mission like Euclid/WFIRST could reduce the variance by a factor of $\sim 0.1$.  

We show the improvement in the $S/N$ for the estimation of $\fnl$ using the RNE assuming 
a satellite experiment like Planck ($\lmax = 2500$, $f_{\rm sky} = 0.8$)  in Fig.~\ref{fig:SN}. If the data indicate that $\fnl \neq 0$, then the estimator discussed here can be used to increase the $S/N$ of the detection, thus shedding new light on primordial non-Gaussianity.

\begin{acknowledgments}
We acknowledge useful conversations with S.~Ho,  D.~N.~Spergel,  B.~Wandelt, O.~Zahn,  and M.~Zaldarriaga. TLS was supported by the Berkeley Center for Cosmological Physics and is grateful for the hospitality of the Institute for the Physics and Mathematics of the Universe (IPMU), University of Tokyo, where part of this work was completed. DG was supported at the Institute for Advanced Study by the National Science Foundation (AST-0807044) and NASA (NNX11AF29G). MK was supported by the Department of Energy (DoE SC-0008108) and NASA (NNX12AE86G).
\end{acknowledgments}

\begin{appendix}
\section{Flat-sky approximation} 

To speed computation throughout the paper (by allowing the use of FFTs along both dimensions of the sky), we use the flat-sky approximation as described in Refs.  \cite{Kamionkowski:2010me,Smith:2011rm}, while providing whole-sky expressions for use on real maps.  In terms of the fractional
temperature perturbation $T(\hat{n})$ at position $\hat{n}$, the temperature power spectrum
$C_l$ is given by
\begin{eqnarray}
     \VEV{a_{\vec l_1} a_{\vec l_2}} &=& \Omega \delta_{\vec
     l_1+\vec l_2,0} C_l,\\
       a_{\vec l} &=& \int\, d^2\vec \theta\, e^{-i\vec l\cdot
     \vec \theta} T(\vec\theta) \simeq
     \frac{\Omega}{N_{\mathrm{pix}}} \sum_{\vec\theta} e^{-i\vec l\cdot
     \vec \theta} T(\vec\theta),
\label{eqn:powerspectrum}
\end{eqnarray}
where $\Omega=4\pi f_{\mathrm{sky}}$ is the survey area (in
steradians), 
and $\delta_{\vec
l_1+\vec l_2,0}$ is a Kronecker delta that sets $\vec l_1 =
-\vec l_2$ and we use the convention $\delta^{2}(\vec{l})=\Omega \delta_{\vec{l}}$ to convert between Kronecker and Dirac-$\delta$ functions. We convert between discrete sums and integrals using the correspondence $\Sigma_{\vec{l}}\leftrightarrow \Omega \int d^{2}\vec{l}/\left(2\pi\right)^{2}$ \cite{Kamionkowski:2010me}.
\label{flatskyappendix}
We use the convention
\begin{equation}
P(k)=A k^{n_{\rm s}-4},\end{equation} and assume a scale-invariant power spectrum ($n_{s}=1$) for the duration of this paper. For a scale-invariant primordial power spectrum in the Sachs-Wolfe approximation, the angular power spectrum
for $T(\vec\theta)$ is \cite{Babich:2004yc}:
\begin{equation}
     C_l = \frac{A}{9\pi l(l+1)},
     \label{eq:Cl}
\end{equation}
where we take the amplitude $A=2\pi^{2} \Delta_{\Phi}^{2}\simeq 2.43 \times 10^{-9}\times 2\pi^{2}\simeq 4.7\times 10^{-8}$ \cite{Komatsu:2010fb}.
In the flat-sky limit, the reduced bispectrum $b_{l_1 l_2 l_3}$ is given by 
\begin{equation}
     \VEV{a_{\vec l_1} a_{\vec l_2} a_{\vec l_3}} = \Omega
     \delta_{\vec l_1 +\vec l_2 +\vec l_3,0} b_{l_1 l_2 l_3}.
\label{eqn:bispectrum}
\end{equation}
The expression for $b_{l_1 l_2 l_3}$ in Eq.~(\ref{eq:redbi}) itself [in terms of $\alpha_{l}(r)$, $\beta_{l}(r)$] is unchanged. 
The Kronecker delta insures that the bispectrum is defined only
for $\vec l_1 +\vec l_2+\vec l_3=0$; i.e., only for triangles in
Fourier space.  Statistical isotropy then dictates that the
bispectrum depends only on the magnitudes $l_1$, $l_2$, $l_3$ of
the three sides of this Fourier triangle.  To derive various expressions in the text, we will use the flat-sky equivalent of the Wigner-3J coefficient \cite{Hu:2000ee}:
\begin{equation}
\sqrt{\frac{\left(2l_{1}+1\right)\left(2l_{2}+1\right)\left(2l_{3}+1\right)}{4\pi}}\wigner{l_{1}}{0}{l_{2}}{0}{l_{3}}{0}\wigner{l_{1}}{m_{1}}{l_{2}}{m_{2}}{l_{3}}{m_{3}}
\to\delta_{\vec{l}_{1}+\vec{l}_{2}+\vec{l}_{3},0}\Omega.\label{komatsu_flatfull_wigner}\end{equation}

In the flat-sky limit, the MVNH is given (applying the same arguments used to derive it in the whole-sky case) by
\cite{Babich:2004yc,Kamionkowski:2010me}
\begin{equation}
     \fnlnull \equiv \sigma_{0}^{2} \sum_{\vec l _1+ \vec l_2+
     \vec l_3=0} \frac{ a_{\vec l_1} a_{\vec
     l_2} a_{\vec l_3}b_{l_1 l_2 l_3}}{ 6\Omega^2 
     C_{l_1}C_{l_2}C_{l_3}},
\label{eqn:biestimator}
\end{equation}
and it has inverse variance,
\begin{equation}
     \sigma_{0}^{-2} = \sum_{\vec l _1+ \vec l_2+ \vec l_3=0} \frac{
     \left[ b_{l_1 l_2 l_3} \right]^2}{6 \Omega
     C_{l_1}C_{l_2}C_{l_3}}.
\label{eqn:binoise}
\end{equation}

\section{Details of the calculation of the variance}
\label{variance_details}
\begin{figure}[ht]
\begin{center}
\resizebox{!}{7cm}{\includegraphics{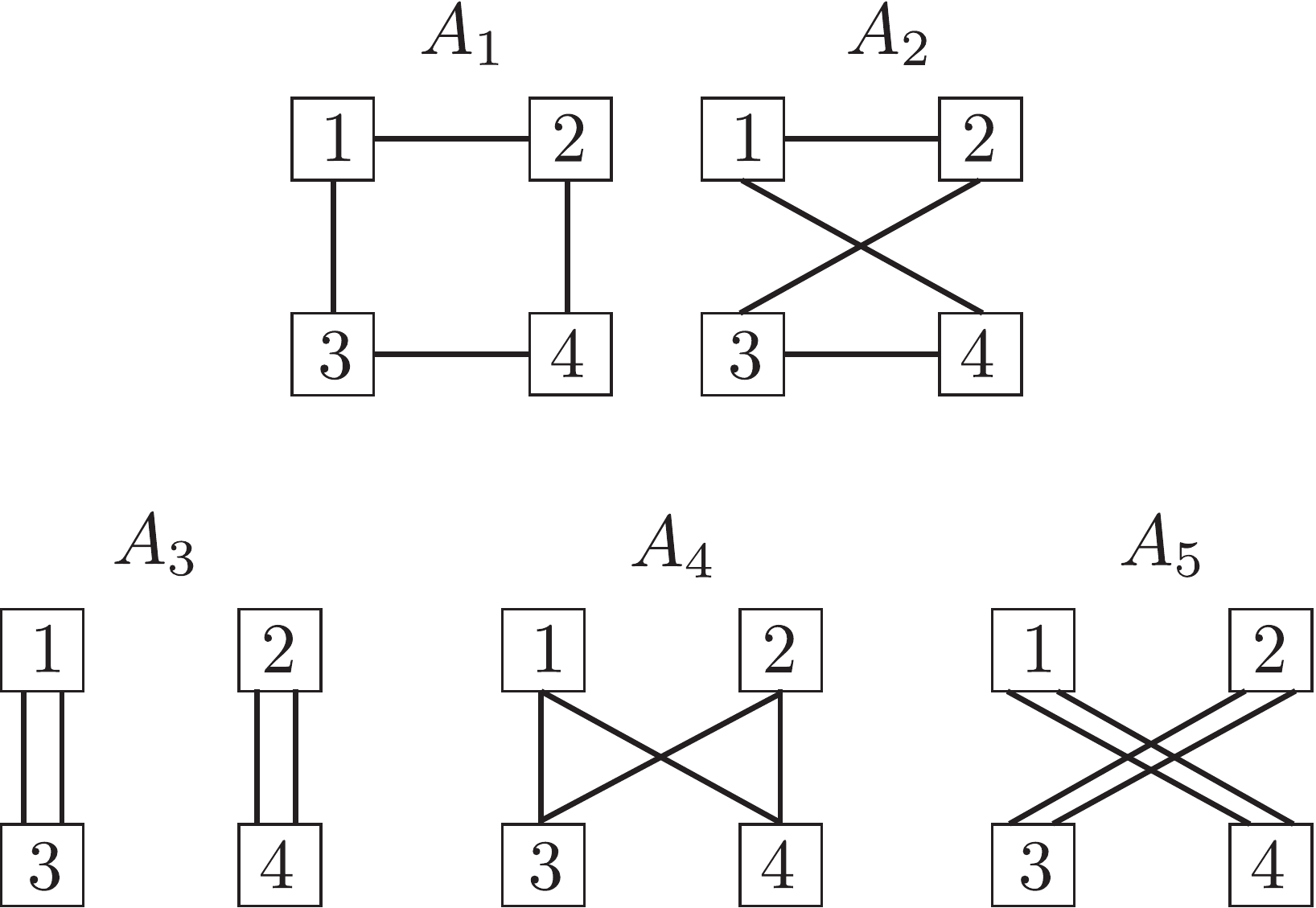}}
\caption{A graphical representation of the 5 combinations of $1 = (\vec l_2, \vec l_3)$, $2 = (\vec k, \vec k')$, $3 = (\vec m_2, \vec m_3)$, 
$4 = (\vec p, \vec p')$ which make up the variance of $\mathcal{B}_1$.}
\label{fig:combb}
\end{center}
\end{figure}

We will first compute the variance of $\mathcal{B}_1$.  To do so we will concentrate on the part of the variance 
\begin{eqnarray}
&& \int r^2 dr (r')^2 dr'  \alpha_{l_1}(r) \alpha_{m_1}(r') \VEV{a_{\vec l_2} a_{\vec l_3} \phi_{\vec k}(r) \phi_{\vec k'}(r)|a^*_{\vec m_2} a^*_{\vec m_3} \phi^*_{\vec p}(r') \phi^*_{\vec p'}(r')}\nonumber \\ &\times& \delta_{\vec l_1 + \vec l_2 + \vec l_3,0} \delta_{\vec k + \vec k',\vec l_1}\delta_{\vec m_1 + \vec m_2 + \vec m_3,0} \delta_{\vec p + \vec p',\vec m_1}.\label{varfac}
\end{eqnarray}
We will first concentrate on identifying the various types of terms that will arise from the expectation value.  First note that the 
Kronecker deltas require that the only `internal' contractions can be between the $\vec l$ or $\vec m$ terms with the $\vec k$s or $\vec p$s.
Therefore the only contractions on the `off-diagonal' [in which all $a_{\vec{l}}$s are contracted with $\phi_{l'}(r)$] are:
\begin{eqnarray}
A_1&=&\VEV{a_{\vec l_2} \phi_{\vec k}(r)} \VEV{a_{\vec m_2}^* \phi^*_{\vec p}(r')} \VEV{a_{\vec l_3} a_{\vec m_3}^*} \VEV{\phi_{\vec k'}(r) 
\phi_{\vec p'}(r')^*} ,\\
&=& \delta_{\vec l_2, -\vec k} \beta_{l_2}(r) \delta_{\vec m_2, -\vec p} \beta_{m_2}(r') 
C_{l_3} \delta_{\vec l_3,\vec m_3}\chi_{k'}(r,r') \delta_{\vec k',\vec p'}, \\
A_2 &=& \VEV{a_{\vec l_2} \phi_{\vec k}(r)} \VEV{a_{\vec m_2}^* \phi^*_{\vec p}(r')}  \VEV{a_{\vec l_3} \phi^*_{\vec p'}(r')} \VEV{\phi_{\vec k'}(r) 
a_{\vec m_3}^*}, \\
&=& \delta_{\vec l_2, -\vec k} \beta_{l_2}(r) \delta_{\vec m_2, -\vec p} \beta_{m_2}(r') 
 \delta_{\vec l_3,\vec p'} \beta_{l_3}(r') \delta_{\vec m_3,\vec k'} \beta_{m_3}(r).
\end{eqnarray}
There are 4!=24 `diagonal' contractions; however, not all are unique since the sum is symmetric in $(\vec l_1$,$\vec l_2)$, 
 $(\vec k$,$\vec k')$, $(\vec m_1$,$\vec m_2)$, $(\vec p$,$\vec p')$.  Representing these pairs by numbered boxes in Fig.~\ref{fig:combb} we 
 show the five unique combinations that will make up the variance.
The other three unique terms are: 
\begin{eqnarray}
A_3&=&\VEV{a_{\vec l_2} a^*_{\vec m_2}}\VEV{a_{\vec l_3} a^*_{\vec m_3}}\VEV{\phi_{\vec k}(r) \phi^*_{\vec p}(r')}\VEV{\phi_{\vec k'}(r)
\phi^*_{\vec p'}(r')}\\ &=& C_{l_2} \delta_{\vec l_2, \vec m_2} C_{l_3}  \delta_{\vec l_3, \vec m_3} \chi_{k}(r,r') \delta_{\vec k,\vec p} \chi_{k'}(r,r')
\delta_{\vec k',\vec p'}.
\end{eqnarray}
\begin{eqnarray}
A_4=&&\VEV{a_{\vec l_2} a^*_{\vec m_2}} \VEV{a_{\vec l_3} \phi^*_{\vec p}(r')} \VEV{\phi_{\vec k}(r) a^*_{\vec m_3}} \VEV{\phi_{\vec k'}(r) \phi^*_{\vec p'}(r)} \\
&=& C_{l_2} \delta_{\vec l_2, \vec m_2} \beta_{l_3}(r') \delta_{\vec l_3, \vec p} \beta_{m_3}(r) \delta_{\vec k, \vec m_3} \chi_{k'}(r,r') \delta_{\vec k',\vec p'}.
\end{eqnarray}
\begin{eqnarray}
A_5 &=&\VEV{a_{\vec l_2} \phi^*_{\vec p}(r')}\VEV{a_{\vec l_3} \phi^*_{\vec p'}(r)}\VEV{\phi_{\vec k}(r) a^*_{\vec m_2}}\VEV{\phi_{\vec k'}(r')a^*_{\vec m_3} } \\
&=& \beta_{l_2}(r') \delta_{\vec l_2, \vec p} \beta_{l_3}(r) \delta_{\vec l_3,\vec p'} \beta_{m_2}(r) \delta_{\vec k,\vec m_2} \beta_{m_3}(r') \delta_{\vec m_3, \vec k'}.
\end{eqnarray}

Therefore, in the end we have five unique combinations:
\begin{eqnarray}
A_1 &=& C_{l_3} \delta_{\vec l_3, \vec m_3} \beta_{l_2}(r) \delta_{\vec l_2, -\vec k} \beta_{m_2}(r') \delta_{\vec m_2, -\vec p} \chi_{k'}(r,r') \delta_{\vec k', \vec p'},\\
A_2 &=& \beta_{l_2}(r) \delta_{\vec l_2, -\vec k} \beta_{m_3}(r) \delta_{\vec m_3, \vec k'} \beta_{l_3}(r') \delta_{\vec l_3, \vec p'} 
\beta_{m_2}(r') \delta_{\vec m_2, -\vec p},\\
A_3 &=& C_{l_2} \delta_{\vec l_2, \vec m_2} C_{l_3} \delta_{\vec l_3, \vec m_3} \chi_{k}(r,r') \delta_{\vec k, \vec p}\chi_{k'}(r,r') \delta_{\vec k', \vec p'},\\
A_4 &=&C_{l_2} \delta_{\vec l_2, \vec m_2} \beta_{m_3}(r) \delta_{\vec m_3, \vec k} \beta_{l_3}(r') \delta_{\vec l_3, \vec p} \chi_{k'}(r,r') \delta_{\vec k', \vec p'},\\
A_5 &=& \beta_{l_3}(r) \delta_{\vec l_3, \vec p'} \beta_{m_2}(r) \delta_{\vec m_2, \vec k} \beta_{l_2}(r') \delta_{\vec l_2, \vec p} 
\beta_{m_3}(r') \delta_{\vec m_3, \vec k'}.
\end{eqnarray}

The last term, $A_5$, has the Kronecker deltas $\delta_{\vec l_3, \vec p'} \delta_{\vec l_2, \vec p}$ which implies the full 
term will have the Kronecker delta $\delta_{\vec l_1 + \vec l_2 + \vec l_3, 0} \delta_{\vec p + \vec p', \vec l_1} \delta_{\vec l_3, \vec p'} \delta_{\vec l_2, \vec p}$ so that when summing over $\vec p$ and $\vec p'$ we will have $\vec l_2 = \vec p$ and $\vec l_3 = \vec p'$ so the 
final term will be zero since $| \vec l_1| \geq 2$.  Therefore $A_5 = 0$ and we are left with four unique terms, which agrees with 
the appendix in Ref.~\cite{Smith:2011rm}. 
The same approach can be taken with the variance of $\widehat{\mathcal{B}}_1$
\begin{equation}
\VEV{(\Delta \widehat{\mathcal{B}}_1)^2} \sim \frac{\beta_k(r) \beta_{k'}(r)}{C_{k} C_{k'}}\frac{\beta_p(r') \beta_{p'}(r')}{C_{p} C_{p'}}\VEV{a_{\vec l_2} a_{\vec l_3} a_{\vec k} a_{\vec k'} | 
a^*_{\vec m_2} a^*_{\vec m_3} a^*_{\vec p} a^*_{\vec p'}},
\end{equation}
and the covariance between $\widehat{\mathcal{B}}_1$ and $\mathcal{B}_1$,
\begin{equation}
\VEV{\Delta \widehat{\mathcal{B}}_1\Delta \mathcal{B}_1}  \sim \frac{\beta_k(r) \beta_{k'}(r)}{C_{k} C_{k'}}\VEV{a_{\vec l_2} a_{\vec l_3} a_{\vec k} a_{\vec k'} | 
a^*_{\vec m_2} a^*_{\vec m_3} \phi^*_{\vec p}(r') \phi^*_{\vec p'}(r')}.
\end{equation}
Computing these terms shows that certain contractions dominate the sum so that \cite{Smith:2011rm}
\begin{eqnarray}
\VEV{\Delta (\mathcal{B}_1)\Delta(\mathcal{B}_1)} &=& 8 \sum_{\{\vec l\},\{\vec k\}} \frac{b_{l_1 l_2 l_3}b_{k_1 k_2 k_3} }{C_{l_1} C_{l_2} C_{k_1} C_{k_2} C_{k_3}}\delta_{\vec l_3, \vec k_3}  \\
&\times& \int r^2 dr (r')^2 dr' \alpha_{l_1}(r) \alpha_{k_1}(r')\beta_{k_2}(r') \beta_{l_2}(r) \chi_{l_3}(r,r') ,\nonumber \\
\VEV{\Delta \widehat{(\mathcal{B}_1)}\Delta \widehat{(\mathcal{B}_1)}} &=&
 8 \sum_{\{\vec l\},\{\vec k\}} \frac{b_{l_1 l_2 l_3} b_{k_1 k_2 k_3} }{C_{l_1} C_{l_2} C_{k_1} C_{k_2} C_{k_3}}\delta_{\vec l_3, \vec k_3}  \\
&\times& \int r^2 dr (r')^2 dr' \alpha_{l_1}(r) \alpha_{k_1}(r')\beta_{k_2}(r') \beta_{l_2}(r) \frac{\beta_{l_3}(r) \beta_{l_3}(r')}{C_{l_3}},\nonumber \\
\VEV{\Delta (\mathcal{B}_1)\Delta \widehat{(\mathcal{B}_1)}}&=& \VEV{\Delta (\mathcal{B}_1)\Delta(\mathcal{B}_1)}.
\end{eqnarray}

\section{Fast algorithm to compute RNE}

As noted in Refs. \cite{Komatsu:2003iq,Munshi:2009ik}, due to the inefficiency of harmonic transforms, the MVNH estimator is expensive to evaluate, requiring the computation of $\sim l_{\rm max}^{5} $ terms. However, a more efficient computational algorithm can be used once the MVNH estimator is written in terms real-space quantities- once this is done, the azimuthal part of the harmonic transform can be computed using computationally efficient fast-Fourier transforms (FFTs).  As is noted in Ref.~\cite{Komatsu:2003iq} the MVNH estimator can be rewritten as 
 \begin{eqnarray}
 \fnle&=& \sigma_{0}^{-2} \int r^{2} dr \int d^{2}\hat{n} B^{2}(\hat{n},r)A(\hat{n},r),\\
 A (\hat{n},r)&=&\sum_{lm}\frac{\alpha_{l}(r)Y_{lm}(\hat{n}) a_{lm}}{C_{l}},\label{adef}\\
 B(\hat{n},r)&=&\sum_{lm}\frac{\beta_{l}(r)Y_{lm}(\hat{n}) a_{lm}}{C_{l}}.\label{bdef}
 \end{eqnarray}
At each location along the line of sight, the resulting estimator only requires the computation of $\sim l_{\rm max}^{3}$ terms.  In addition to this, the filter functions $\alpha_{l}(r)$ and $\beta_{l}(r)$ are sufficiently smooth so that they must only be evaluated for $\mathcal{O}(100)$ grid points.  In Sec. \ref{rdn}, we generalized the realization-dependent normalization of Ref.~\cite{Creminelli:2006gc} to treat the full sky and include the radiation transfer function. Here we show how we may rewrite this estimator in order to utilize FFTs to speed up their computation.  

The estimator is given by 
\begin{equation}
\widehat{\mathcal{B}_1} =\sigma_{0}^{-2}\sum_{l\leq l_{1}\leq l_{2},l_{a}l_{b}}\sum_{m_{1}m_{2}m_{a}m_{b}m}
     \frac{a_{l_1 m_1}a_{l_2 m_2}a_{l_{a}m_{a}}a_{l_{b}m_{b}}}{2C_{l} C_{l_1} C_{l_2}C_{l_{a}}C_{l_{b}} }B_{ll_1l_2}B_{ll_{a}l_{b}}\wigner{l}{m}{l_{1}}{m_{1}}{l_{2}}{m_{2}}\wigner{l}{m}{l_{a}}{m_{a}}{l_{b}}{m_{b}}.\label{estappend}
     \end{equation} Using Eq.~(\ref{gaunt}), (\ref{adef}), and (\ref{bdef}), we may rewrite Eq.~(\ref{estappend}) in a form amenable to rapid computation using filtered real-space maps and FFTs:
\begin{eqnarray}
\mathcal{\hat{B}}_{1}&=&\sigma_{0}^{-2}\sum_{lm} \left\{\frac{\mathcal{V}_{lm}\mathcal{V}_{lm}^{*}+4U_{lm}U_{lm}^{*}+2\left(U_{lm}\mathcal{V}_{lm}^{*}+U_{lm}^{*}\mathcal{V}_{lm}\right)}{2C_{l}}\right\}\label{vint_tosave},\\
\mathcal{V}_{lm}&=&\int dr r^{2} \alpha_{l}(r)B_{lm}^{(2)}(r),\label{vintdef}~~B_{l m}^{(2)}(r)=\int d\hat{n} Y_{lm}^{*}(\hat{n}) B^{2}(r,\hat{n}),\\
U_{lm}&=&\int dr r^{2} \beta_{l}(r) \int d\hat{n}Y_{lm}^{*}(\hat{n})A(r,\hat{n})B(r,\hat{n})\label{uintdef}.
\end{eqnarray}
Written this way, the normalization of the estimator may be computed from a map, and may be efficiently evaluated using FFTs, although the computation of $\mathcal{V}_{lm}$ via numerical integration does introduce a new bottleneck. The operation count for these procedures is $\sim l_{\rm max}^{5/2} \ln{(l_{\rm max})}N_{\rm int}$ where $N_{\rm int}$ is the number of points sampled along the line of sight for the radial integral. The $l_{\rm max}^{5/2}\ln{(l_{\rm max})}$ scaling follows from the computational expense of a harmonic transform (in which an FFT is used for the azimuthal Fourier transform piece). In contrast, the direct evaluation of Eq.~(\ref{eq:B1estimator_a}) would require evaluating and summing $\sim l_{\rm max}^{10}$ terms. The computational savings is then a factor of $l_{\rm max}^{7.5}/(\ln{(l_{\rm max})}N_{\rm int})$, a huge savings for $\lmax \sim 10^{3}$, as is the case for the WMAP and Planck missions. In Ref.~\cite{Komatsu:2003iq}, it is found that to obtain convergence in the bispectrum estimator itself, $N_{\rm int}\sim 10^{3}$ is more than adequate. We thus estimate that use of the real-space RNE will to a computational savings factor of $10^{18}$ over the harmonic-space estimator.

\end{appendix}
\bibliography{paper}

\end{document}